\documentclass[journal,comsoc]{IEEEtran}

\usepackage[T1]{fontenc}
\usepackage{booktabs} 
\usepackage{amssymb}
\usepackage{amsthm}

\usepackage{amsmath}

\usepackage{colortbl}

\usepackage[figuresright]{rotating}

\usepackage{lineno,hyperref}

\usepackage{xcolor}
\definecolor{mycyan}{gray}{.8}

\usepackage{setspace}

\usepackage{cite}
\usepackage[numbers,sort&compress]{natbib}
\usepackage{amsmath}
\usepackage[cmintegrals]{newtxmath}

\usepackage{booktabs}
\usepackage{multirow}
\usepackage{supertabular}
\usepackage{gensymb}

\usepackage{threeparttable}

\usepackage{makecell}

\usepackage{flushend}
\usepackage[linesnumbered,ruled]{algorithm2e}
\SetKwProg{Fn}{Function}{}{end}\SetKwFunction{FRecurs}{FnRecursive}%

\usepackage{ulem}

\usepackage{soul}
\soulregister\cite7 
\soulregister\citep7 
\soulregister\citet7 
\soulregister\ref7 
\soulregister\pageref7 

\ifCLASSOPTIONcompsoc
\usepackage[caption=false,font=normalsize,labelfont=sf,textfont=sf]{subfig}
\else
\usepackage[caption=false,font=footnotesize]{subfig}
\fi

\begin{document}
	
	\title{Genetic U-Net: Automatically Designed Deep Networks for Retinal Vessel Segmentation Using a Genetic Algorithm}
	
\author{
	Jiahong Wei \qquad Zhun Fan\\
	Department of Electronic Engineering, Shantou University, Guangdong, 515063, China\\
	Key Lab of Digital Signal and Image Processing of Guangdong Province, Guangdong, 515063, China\\
	Email: 19jhwei@stu.edu.cn \quad zfan@stu.edu.cn\\
	\thanks{Corresponding author: Zhun Fan}
}

\maketitle

\begin{abstract}
Recently, many methods based on hand-designed convolutional neural networks (CNNs) have achieved promising results in automatic retinal vessel segmentation. However, these CNNs remain constrained in capturing retinal vessels in complex fundus images. To improve their segmentation performance, these CNNs tend to have many parameters, which may lead to overfitting and high computational complexity. Moreover, the manual design of competitive CNNs is time-consuming and requires extensive empirical knowledge. Herein, a novel automated design method, called Genetic U-Net, is proposed to generate a U-shaped CNN that can achieve better retinal vessel segmentation but with fewer architecture-based parameters, thereby addressing the above issues. First, we devised a condensed but flexible search space based on a U-shaped encoder-decoder.
Then, we used an improved genetic algorithm to identify better-performing architectures in the search space and investigated the possibility of finding a superior network architecture with fewer parameters.
The experimental results show that the architecture obtained using the proposed method offered a superior performance with less than 1\% of the number of the original U-Net parameters in particular and with significantly fewer parameters than other state-of-the-art models. Furthermore, through in-depth investigation of the experimental results, several effective operations and patterns of networks to generate superior retinal vessel segmentations were identified.
\end{abstract}


\IEEEpeerreviewmaketitle

\section{Introduction}
\label{intro}
The retinal vascular system can be observed non-invasively in vivo in humans \cite{vostatek2017performance}. In addition, changes in the vasculature are often associated with certain diseases, leading ophthalmologists and other physicians to consider fundus examination a routine clinical examination \cite{fraz2012blood,chatziralli2012value}. Many diseases can be diagnosed and tracked \cite{fraz2012blood} by observing the retinal vascular system. Pathological changes in retinal vessels can reflect either ophthalmology diseases or other systemic diseases, such as wet age-related macular degeneration and diabetes \cite{abramoff2010retinal}. Diabetic retinopathy can lead to the growth of new blood vessels, and atherosclerosis \cite{hubbard1999methods} associated with wet age-related macular degeneration can cause the narrowing of blood vessels. 
Moreover, the retinal vascular system of each eye is unique. Without pathological changes, it does not alter throughout the lifetime. Hence, observation of the retinal vascular system can also be applied in biometrics \cite{ortega2009personal,simon1935new}. Through retinal vessel segmentation, relevant morphological information of retinal vascular trees (such as the width, length, and curvature of blood vessels) can be obtained \cite{jin2019dunet}. 
Consequently, precise retinal vessel segmentation is significant. However, owing to the complexities of retinal vascular structures, manual inspection is subjective, time-consuming, and laborious \cite{mou2019dense,wang2019dual}. Therefore, developing an effective algorithm for the automated segmentation of retinal vessels to support ophthalmologists in clinical assessment has been of great interest.

Due to the complexities of fundus images, automated segmentation of retinal vessels is challenging. First, in fundus images, retinal vessels are difficult to distinguish from the background because of the subtle difference between the vascular region and background. Second, the structures of vascular trees are also complicated, with many cross-connected and minuscule vessels. Third, other factors such as pathological exudates and uneven illumination render segmentation difficult. Compared with the methods \cite{staal2004ridge,hoover2000locating,odstrcilik2013retinal,orlando2016discriminatively,fan2016automated_1} based on traditional image processing or handcrafted features, the deep convolutional neural network (DNN) methods \cite{liskowski2016segmenting,vega2015retinal,li2015cross,mo2017multi,fan2016automated_2} offer advantages in dealing with these complications.
In particular, U-Net \cite{ronneberger2015u} and its variants \cite{jin2019dunet,wu2019vessel,wang2019dual,mou2019dense,laibacher2019m2u,alom2019recurrent,gu2019net,mou2021cs2} have recently become the mainstream models used in retinal vessel segmentation.

However, these U-Net variants are designed manually and have two main shortcomings that constrain them in clinical applications. First, they still have deficiencies in extracting the features of vascular trees from complicated fundus images, which limits their capability of dealing with challenging cases (e.g., lesions areas, low contrast micro-vessels). Second, to improve segmentation performance, these manually designed U-Net variants tend to have many parameters. However, because constructing vessel segmentation datasets is laborious and highly dependent on expertise, only a small amount of annotated data is available. A network architecture with too many parameters is prone to overfitting when using insufficient training data and usually has high computational complexity. Moreover, the original U-Net and its variants utilize one or two types of identical blocks to form the architecture. However, different blocks in a U-shaped network are required to process different patterns of features in fundus images. Therefore, having the same structures for different blocks may not be optimal, as it limits the feature extraction ability and parameter efficiency \cite{huang2017densely} of the U-shaped networks.

Conversely, designing improved network architectures manually requires rich domain-specific knowledge and considerable time, especially when only a small number of parameters is expected. Therefore, this study developed a Neural Architecture Search (NAS)-based approach to automatically discover network architectures tailored for retinal vessel segmentation. Compared with natural images, fundus images have unique features \cite{mou2021cs2}; therefore, the success of NAS in natural image segmentation \cite{liu2019auto} is not immediately transferable to retinal vessel segmentation. Several attempts have been made to deploy NAS in 2D medical image segmentation \cite{weng2019unet,mortazi2018automatically}, but the search spaces of these methods are highly constrained and usually include the repeated stacking of identical blocks with no topology optimization. The generated model retains many parameters, which renders it unsuitable for retinal vessel segmentation, which, with only a small amount of labeled data available is much more sensitive to overfitting. For NAS, the search space designed by human experts is crucial for the performance of the discovered architectures \cite{fang2020densely,yu2020c2fnas}.

In the proposed Genetic U-Net, we first define a specific search space and then realize the automated design of network architectures with an improved genetic algorithm (GA). When designing the search space, two key aspects are considered 1) defining a condensed but flexible search space, and 2) restricting the total number of architecture-based parameters. Since it is based on a U-Net backbone architecture, the search space is condensed to prevent the algorithm from having to search in an open-ended search space. It is flexible because different blocks of the U-Net architecture are diversified and optimized individually. As diverse optimized internal block topologies can be found in the overall architecture of U-Net, a model with superior performance yet far fewer parameters can be identified. Restricting the total number of architecture-based parameters of the possible architectures in the search space can reduce the computational requirements (e.g., GPU memory) and facilitate the acquisition of high-performance network architectures with even fewer parameters. Moreover, to improve the search efficiency of NAS, an improved GA is suggested with enhanced crossover and selection operations.

The importance of this research is described as follows: First, to our knowledge, this is the first instance of evolutionary NAS being applied to retinal vessel segmentation. Second, the proposed method exploits the potential of the U-shaped encoder-decoder structure and investigates whether an exceptional DNN can be designed with very few parameters and without using complex mechanisms (e.g., multiple encoders, multiple decoders, cascade structures, attention mechanisms). Finally, compared with previous CNNs that have been applied to retinal vessel segmentation, the proposed method can reduce the multiplicity of redundant parameters, which is important when training data are limited.
Extensive experiments were performed to show that Genetic U-Net can design compact architectures that perform better than state-of-the-art models.
The main contributions of the work are summarized below:

\begin{itemize}
	\item We propose a novel automated design method for the U-shaped CNN architecture based on a specific search space and an improved GA, which enables us to acquire compact network architectures that outperform existing ones in retinal vessel segmentation.
	\item Through observation and analysis of the discovered architectures, we observed that some patterns and operations can significantly improve the performance of retinal vessel segmentation. The findings may lead to new insights into the design of network architecture.
	\item Compared with state-of-the-art models, the found models offered superior performance on several public datasets with the least parameters (less than 1\% of the original U-Net).
\end{itemize}

\section{{Related} Work}
\label{sec:2}
\subsection{Retinal Vessel Segmentation}
\label{sec:2.1}

Originally, researchers used traditional image processing techniques to segment retinal vessels, such as thresholding segmentation \cite{hoover2000locating} or certain morphological operations \cite{staal2004ridge,odstrcilik2013retinal}. Under different circumstances, many of these methods’ hyperparameters had to be re-adjusted to achieve satisfactory segmentation results. Later, several learning-based methods \cite{orlando2016discriminatively,fan2016automated_1} incorporating handcrafted features were applied to this task. Since these handcrafted features do not have sufficient generalization ability to represent characteristics in a variety of complicated fundus images, these methods are misled by extreme cases (e.g., lesions areas and low contrast microvessels). Compared with the above methods, the DNNs-based methods \cite{liskowski2016segmenting,vega2015retinal,li2015cross,mo2017multi,fan2016automated_2} demonstrate particular advantages in dealing with complexity in fundus images, where the features are learned directly from the training data.

New state-of-the-art methods \cite{yan2018joint,jin2019dunet,wu2019vessel,wang2019dual,mou2019dense,laibacher2019m2u,alom2019recurrent,gu2019net,mou2021cs2} for retinal vessel segmentation are dominated by deep learning models, especially U-Net variants.
Yan \textit{et al.} \cite{yan2018joint} adopted a joint loss to provide supervision information for U-Net, with two parts responsible for pixel-wise loss and segment-level loss. The joint loss can improve the capability of the model to balance the segmentation between thick and thin vessels.
To better capture the diverse morphologies of vascular trees, Jin \textit{et al.} \cite{jin2019dunet} proposed DU-Net that replaces traditional convolution with deformable convolution.
Wu \textit{et al.} \cite{wu2019vessel} designed a novel inception-residual block for a U-shape network and introduced four supervision paths with different convolution kernel sizes to utilize multi-scale features. 
Alom \textit{et al.} \cite{alom2019recurrent} proposed R2U-Net with a recursive residual layer based on cyclical convolutions to capture features.
Mou \textit{et al.} \cite{mou2019dense} embedded dense dilated convolutional blocks between encoder and decoder cells at the same levels of a U-shape network and used a regularized walk algorithm to post-process model predictions. 
Dual U-Net proposed by Wang \textit{et al.} \cite{wang2019dual} had two encoders. One encoder path was for extracting spatial information and the other was for extracting context information. A novel module was also proposed to combine the information of the two paths. 
Laibacher \textit{et al.} \cite{laibacher2019m2u} utilized pre-trained components of MobileNetV2 \cite{sandler2018mobilenetv2} on imageNet \cite{russakovsky2015imagenet} as the encoder of U-shape network and introduced novel contractive bottleneck blocks for the decoder, which achieved better performance, less computational cost, and faster inference speed.
Gu \textit{et al.} \cite{gu2019net} proposed CE-Net by adopting multiple convolution branches with different receptive fields to capture more high-level information and preserve spatial information for segmentation.
Mou \textit{et al.} \cite{mou2021cs2} included a self-attention mechanism in the U-shape encoder-decoder to improve the hierarchical representation capture ability of the model.
The abovementioned networks, which are unusually complex and have many parameters, were manually designed. To increase the feature extraction capability, some networks must be pre-trained with external datasets. In this study, we automatically design the network architectures that can perform well in extracting features from fundus images but require fewer parameters and no pre-training. 

\subsection{Neural Architecture Search}
\label{sec:2.2}

Neural architecture search (NAS) is an effective technique that assists end-users to design effective deep networks automatically.
At present, depending on the optimization methods used, the three main categories of NAS are: (1) The methods based on reinforcement learning \cite{zoph2018learning,zoph2016neural,baker2016designing} that formulate NAS as a Markov decision process. A controller is used to sample the architecture and learn to generate improved architectures from a process of continuous trial and error; (2) The methods based on evolutionary algorithms \cite{xie2017genetic,lu2019nsga} that formulate NAS as an optimization problem and encode the architectures. Increasingly competitive architectures are generated by applying some genetic operations (e.g., crossover, and mutation) and will be retained as offspring in the next generation. The architectures are optimized from generation to generation until those with satisfactory performance are obtained; (3) For differentiable neural architecture search \cite{liu2018darts,brock2018smash}, each operation option is assigned a weight coefficient. The parameter weights of the architecture and the weights of the operation options are optimized alternately by gradient descent.
NAS has achieved great success in the natural image and medical image analysis \cite{zhu2019v,yang2019searching,liu2019deep,mortazi2018automatically,weng2019unet,kim2019scalable}, including some works that apply NAS to medical image segmentation. In the works of \cite{zhu2019v,yang2019searching,mortazi2018automatically}, for medical image segmentation, the hyperparameters and operations of each layer of the building blocks were optimized, but the topology of the block was relatively fixed. Additionally, in the works \cite{weng2019unet,kim2019scalable}, the structure and operations of one or two types of building blocks were optimized, and then the architecture was constructed by repeatedly stacking them. However, in our work, the topology and operations of each block could be different yet simultaneously be optimized flexibly.

\subsection{Genetic Algorithms}
Genetic Algorithms (GAs) \cite{holland1992adaptation}, one of the evolutionary algorithms, are metaheuristics inspired by evolution. They are widely used to find high-quality solutions for optimization problems (e.g., TSP \cite{grefenstette1985genetic}, neuroevolution \cite{stanley2002evolving,yao1999evolving}, NAS \cite{xie2017genetic}) by executing operators such as mutation, crossover, and selection.
In addition, GAs have been applied to many medical image analysis tasks, such as segmentation \cite{fan2002volumetric}, registration \cite{matsopoulos2004multimodal}, disease detection \cite{quellec2008optimal}, and image denoising \cite{liu2019deep}. Fan \textit{et al.} \cite{fan2002volumetric} used GA to overcome numerical instability of the active model-based volumetric segmentation of brain images. Matsopoulos \textit{et al.} \cite{matsopoulos2004multimodal} employed GA to optimize the parameters of affine transform to better register multimodal retinal images. Quellec \textit{et al.} \cite{quellec2008optimal} adopted GA to optimize the parameters of wavelet transform for the detection of micro-aneurysms in retina photographs. Recently, Liu \textit{et al.} \cite{liu2019deep} utilized GA to find optimal hyperparameters and CNN architectures for medical image denoising. However, to our knowledge, there are still no works on applying GAs to optimize CNN architectures for vessel segmentation in medical image analysis.

\section{The Proposed Method}
\label{sec:3}
In this section, we present the proposed method in detail. We first introduce the search space of the architectures, then explain the method of encoding an architecture into a binary string, and finally explain the genetic algorithm with evolutionary operations (e.g., crossover, mutation, and selection) searching for competitive architectures.

\subsection{The search space and encoding}
\label{sec:3.1}
\subsubsection{Backbone of the Search Space}
\label{sec:3.1.1}
\begin{figure}[htbp]
	\centering
	\includegraphics[width=9cm]{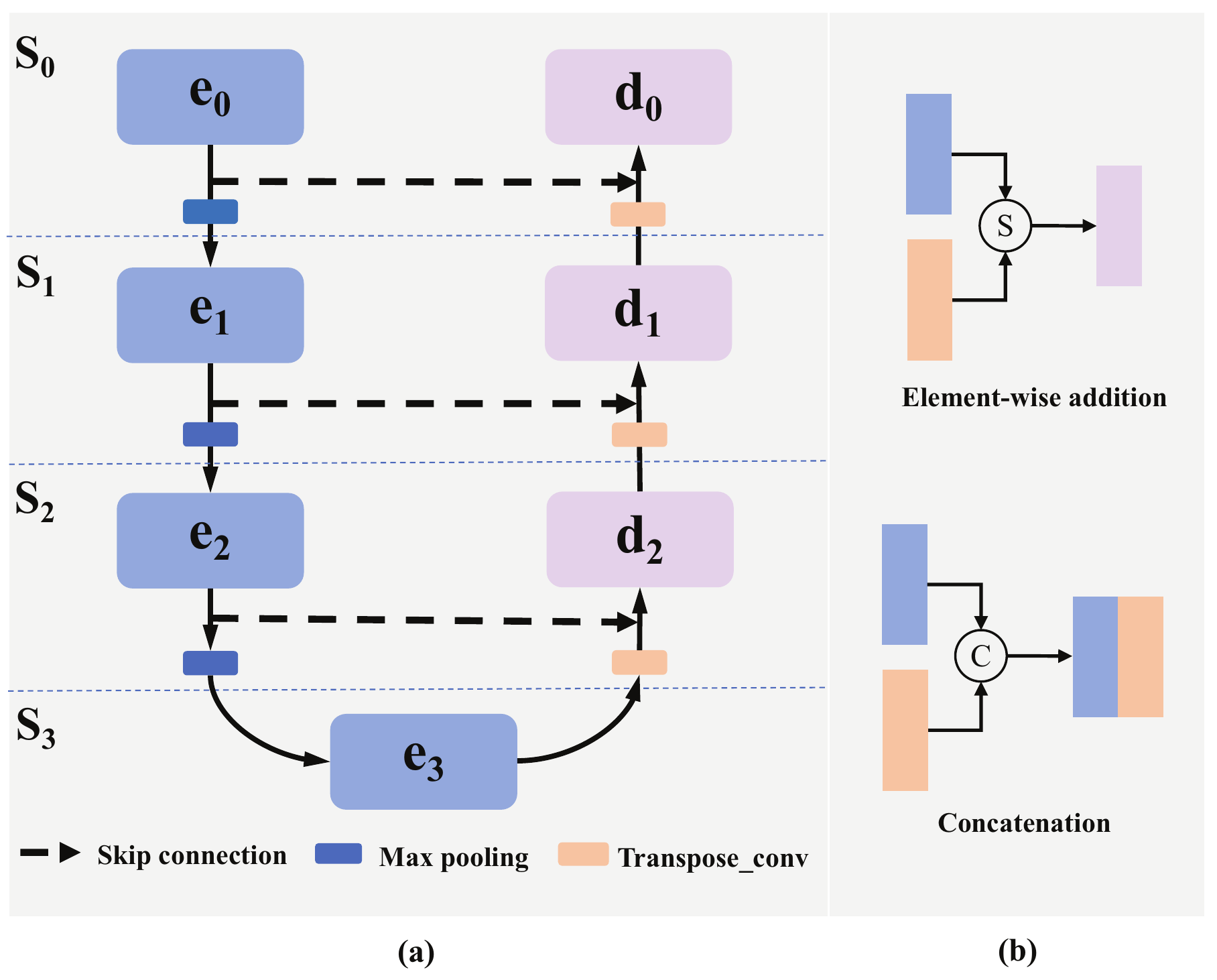}
	\caption{(a) The backbone of U-Net. It is composed of an encoder $E$ and a decoder $D$ and has four stages $S_{0}$, $S_{1}$, $S_{2}$, and $S_{3}$. Encoder $E$ has four building blocks ($e_{0}$, $e_{1}$, $e_{2}$, and $e_{3}$). Decoder $D$ contains three building blocks ($d_{0}$, $d_{1}$, and $d_{2}$); (b) The operations for feature fusion, element-wise addition, and concatenation.}\label{fig:backbone}
	
\end{figure}

As shown in Fig. \ref{fig:backbone}(a), U-Net is composed of an encoder $E$ and a decoder $D$. Both encoder $E$ and decoder $D$ contain several blocks, such as $e_{i}$ $(i=0,1,2,3)$ and $d_{j}$ $(j=0,1,2)$. From top to bottom, U-Net is divided into different stages $S_{k}$ $(k=0,1,2,3)$, and the feature dimensions are constant at each stage.
Skip connections are adopted in all, except for the last, stages to provide features with different semantic information extracted by the encoder to the decoder, which strengthens the connections between the encoder and the decoder and alleviates the vanishing gradient problem \cite{glorot2010understanding}\cite{bengio1994learning} in model training.
Decoder $D$ must fuse features from the skip connections and up-samplings, and the two commonly used feature fusion operations are: concatenation or element-wise addition. Although the original U-Net employs concatenation for feature fusion, some U-Net variants \cite{wang2020non} achieve good results using element-wise addition. Fig. \ref{fig:backbone}(b) illustrates their main differences. Compared with element-wise addition, concatenation generates larger feature maps, which increase computational complexity. To mitigate such complexity in this study, we selected the addition operation for feature fusion.

The successful applications of U-Net and its variants reveal that the U-shaped encoder-decoder structure is highly applicable. Therefore, herein, we used the U-shaped encoder-decoder structure as the backbone. As shown in Fig. \ref{fig:backbone}(a), it consists of seven blocks and four stages. Using this structure, satisfactory architectures can be found by adjusting the internal structures of these building blocks.

\subsubsection{The Building Blocks and Their Encoding}
\label{sec:3.1.2}
In original U-Net, the internal structure of each block is composed of two basic layers ($3\times3$ conv + ReLU). Usually, the U-Net variants \cite{guan2019fully,jegou2017one,gu2019net,alom2019recurrent} improve their performance by adjusting the internal structures of blocks (e.g., ResNet block \cite{He_2016_CVPR}, DenseNet block \cite{huang2017densely} and InceptionNet block \cite{szegedy2015going}), which illustrates the importance of the internal structures of blocks.

The internal structures of the building blocks are represented similarly to Genetic CNN \cite{xie2017genetic}, which is sufficiently flexible to represent many network styles, even including those that are well-known, such as VGG \cite{simonyan2014very}, ResNet \cite{He_2016_CVPR}, and DenseNet \cite{huang2017densely}. Each block can be regarded as a directed acyclic graph consisting of edges and nodes. Each node represents an operation unit or an operation sequence, and the edges represent the connections between nodes. A directed edge between two nodes transforms the output feature map of the pre-node to the post-node. Because the dimensions of all feature maps inside a block are set to be the same, a mismatch would not happen. If a node has more than one edge as the input, the feature maps from these edges will be summed by its elements.

\begin{figure}[htbp]
	\centering
	\includegraphics[width=8cm]{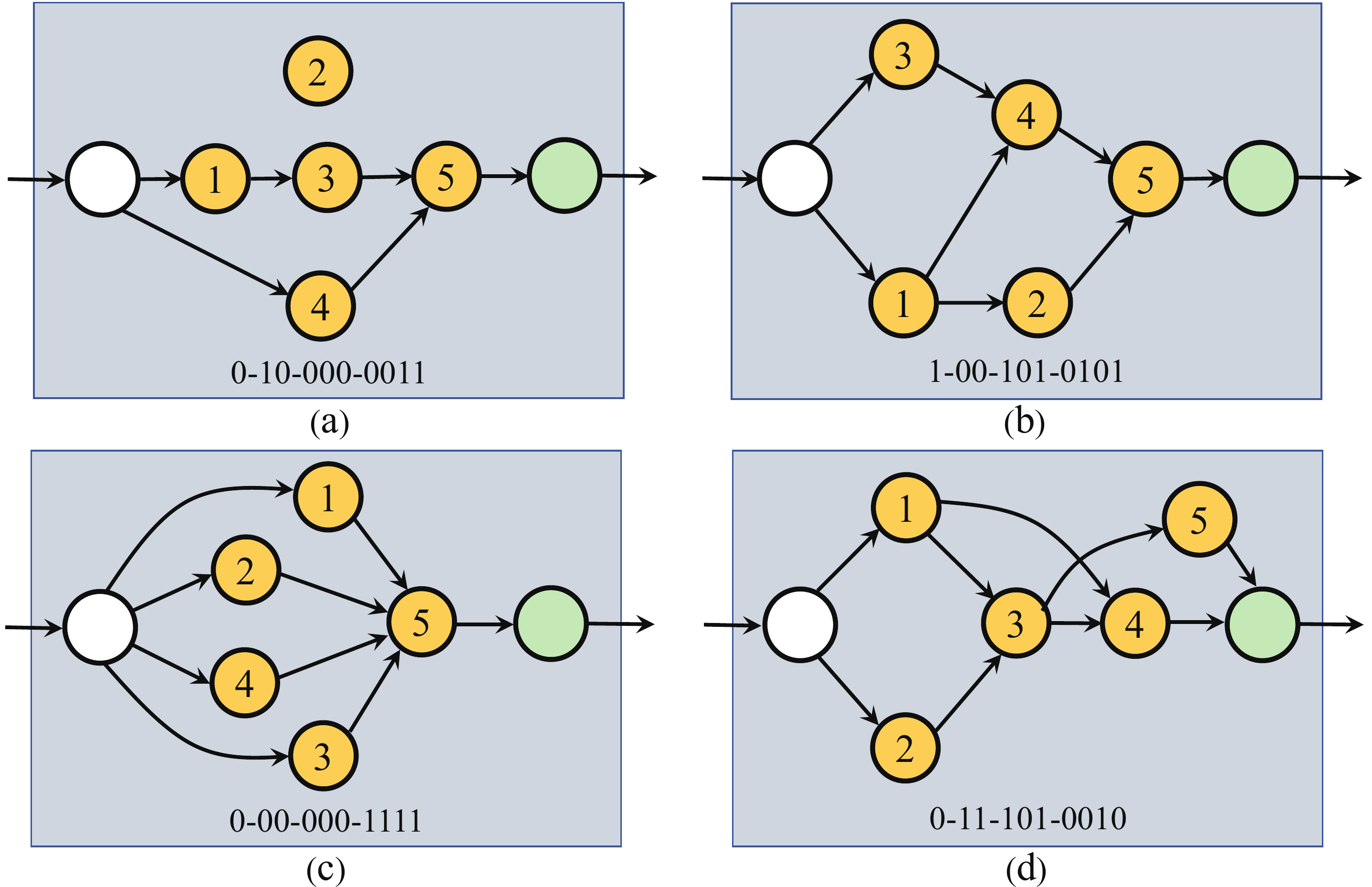}
	\caption{Four encoding examples of the inter-node connections in a block. The white node, the green node, and the yellow nodes represent the default input node, the default output node, and the intermediate nodes ($node_1, node_2,...,node_K$ ($K=5$ in these four examples)). The numbers in the intermediate nodes indicate their orders for encoding. There is a possible maximum of seven nodes in a block in these four examples, and binary encoding with 10 bits represents the connections between nodes in a block. See Section \ref{sec:3.1.2} for detailed descriptions of the encoding schemes of the inter-nodal connections.}\label{fig:encoding}
	
\end{figure}

Fig. \ref{fig:encoding} shows four examples of connections between nodes in a block, and a string with binary encoding represents the inter-node connections.
Assuming the maximum allowed number of intermediate nodes is $K$, then $1+2+3+...+(K-1)=\frac{K(K-1)}{2}$ bits are employed to encode the inter-node connections. The first bit stands for the connection between ($node_1$, $node_2$), the following two represent the connection between ($node_1$, $node_3$) and ($node_2$, $node_3$), etc. If the corresponding bit is 1, the two nodes are connected. For valid encoding, the default input node (marked white in Fig. \ref{fig:encoding}) connects to the intermediate nodes with a successor but without a predecessor. The default output node (marked green in Fig. \ref{fig:encoding}) connects to the intermediate nodes with a predecessor but without a successor. An intermediate node without both a predecessor and a successor, (such as $node_2$ in Fig. \ref{fig:encoding}(a)) is moved out of network architecture, meaning that the number of valid nodes in a block is not fixed.

\begin{figure}[htbp]
	\centering
	\includegraphics[width=8cm]{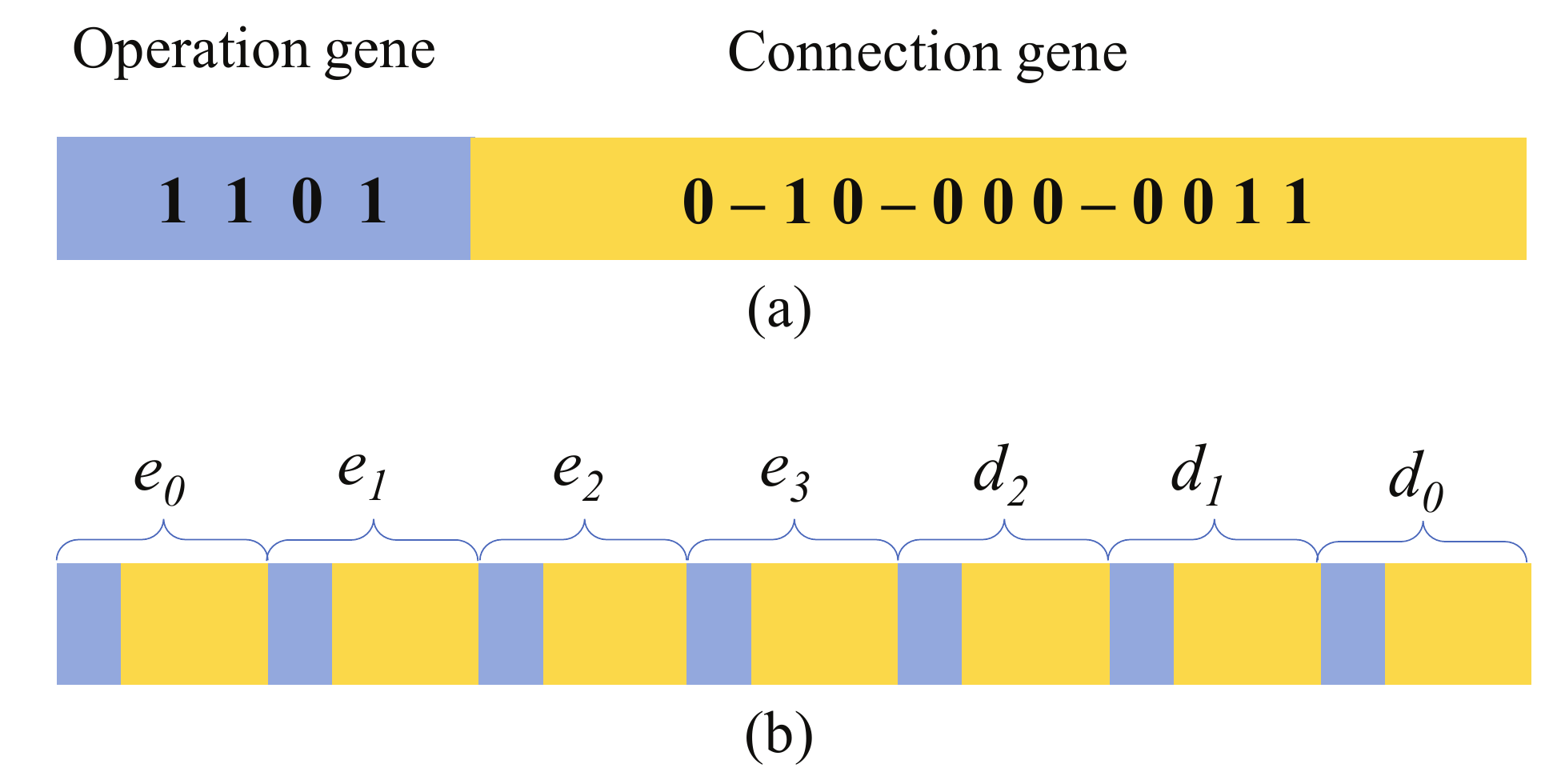}
	\caption{(a) A block gene; (b) Genotype of an architecture consisting of seven block genes.}\label{fig:gene}
	
\end{figure}

All nodes in Genetic CNN \cite{xie2017genetic} have a fixed operational sequence ($3\times3$ conv + BN \cite{ioffe2015batch} + ReLU). In this work, the sixteen operational sequences shown in Table \ref{tab:operations} are provided as options for the nodes. We simultaneously search for the optimal structures and operations of the blocks. Each operational sequence has a unique ID and consists of some basic operational units, including $3\times3$ conv, $5\times5$ conv, ReLU \cite{inproceedings}, Mish \cite{misra2019mish} and instance normalization (IN) \cite{ulyanov2016instance}. ReLU is a generic activation function that performs well on many tasks and is highly applicable. Mish, a relatively recently proposed activation function with similarities to ReLU, also adds continuous differentiability, non-monotonic and other properties, and demonstrates superiority in many tasks. Therefore, both are included in the search space. The above operation units are some commonly used CNN operations, and our goal is to find the most effective sequences of operations for retinal vessel segmentation. The differences between these operational sequences are reflected by the convolutional kernel size, activation functions, activation types (pre-activation or post-activation), and normalization types (e.g., whether instance normalization is utilized). Therefore, four-bit binary encoding is utilized to represent the operational sequences shown in Table \ref{tab:operations}. We assume that each node in one block has the same operation sequence; therefore, each block gene consists of an operation gene with four bits (to represent sixteen different operational sequences) and a connection gene (as shown in Fig. \ref{fig:gene}(a)). Overall, seven block genes together constitute the genotype of architecture (as shown in Fig. \ref{fig:gene}(b)), which indicates that the different blocks can have diverse structures and operations. Therefore, the designed search space is highly flexible.

\begin{table}[htbp]
	\centering
	\caption{The operational sequences for the nodes.}
	\scalebox{0.8}[0.8]{
		\begin{threeparttable}
		\begin{tabular}{clcl}
			\toprule[1.5pt]
			ID    & Operation sequence & ID    & Operation sequence \\
			\midrule[1.5pt]
			0     & $3 \times 3$ conv $\rightarrow $ ReLU & 8     &  ReLU $\rightarrow 3\times3$ conv \\
			1     & $3 \times 3$ conv $\rightarrow $ Mish & 9     & Mish $\rightarrow 3\times3$ conv \\
			2     & $3\times3$ conv $\rightarrow$ IN $\rightarrow$ ReLU & 10    & IN $\rightarrow$ ReLU $\rightarrow 3\times 3$ conv \\
			3     & $3\times3$ conv $\rightarrow$ IN $\rightarrow$ Mish & 11    & IN $\rightarrow$ Mish $\rightarrow 3\times 3$ conv \\
			4     & $5 \times 5$ conv $\rightarrow $ ReLU & 12    & ReLU $\rightarrow 5\times5$ conv \\
			5     & $5 \times 5$ conv $ \rightarrow $ Mish & 13    & Mish $\rightarrow 5\times5$ conv \\
			6     & $5\times5$ conv $ \rightarrow$ IN $\rightarrow$ ReLU & 14    & IN $\rightarrow$ ReLU $\rightarrow 5\times 5$ conv \\
			7     & $5\times5$ conv $\rightarrow$ IN $\rightarrow$ Mish & 15    & IN $\rightarrow$ Mish $\rightarrow 5\times 5$ conv \\
			\bottomrule[1.5pt]
		\end{tabular}%
	\label{tab:operations}%
\begin{tablenotes}
\footnotesize
\item IN represents the instance normalization, ReLU indicates Rectified Linear Unit, and Mish is a self-regularized non-monotonic neural activation function.
\end{tablenotes}
\end{threeparttable} 
}
\end{table}%

The proposed approach has four fundamental differences from Genetic CNN \cite{xie2017genetic}. First, the studied task is different. Genetic CNN is applied to natural image classification, while the proposed method is designed for retinal vessel segmentation. Second, the search space is different. The U-shaped encoder-decoder structure is employed as the backbone in our work but not utilized in Genetic CNN. In addition, the operations of the nodes can be optimized instead of a fixed operation in Genetic CNN. Third, the number of architecture-based parameters in the search space, herein, is constrained within a range of small values dedicated to search for architectures with fewer parameters. Last, an improved GA is proposed to promote search efficiency.

\subsubsection{ Number of Architecture-based Parameters}
The number of the architecture parameters in the search space is closely related to the number of stages in the U-shape backbone, number of intermediate nodes in each block, and number of channels for convolution operations in the nodes. After selecting the backbone adopted for the search space to constrain the number of the architecture parameters and maintain architectural flexibility, the maximum allowed number of intermediate nodes in each block and the number of channels for convolution operations in the nodes are set to 5 and 20, relatively compromised or small values, respectively. Most network architectures in this search space have fewer than 0.4 Million parameters, while normal architectures used for retinal vessel segmentation have more than several Million or even dozens of Million parameters. By restricting the number of architecture-based parameters of the possible architectures within a range of smaller values allows better architectures with greater parameterization efficiency to be found.

\subsection{Evolutionary Algorithm}
\label{sec:3.2}
\begin{figure*}[htbp]
	\centering
	\includegraphics[width=16cm]{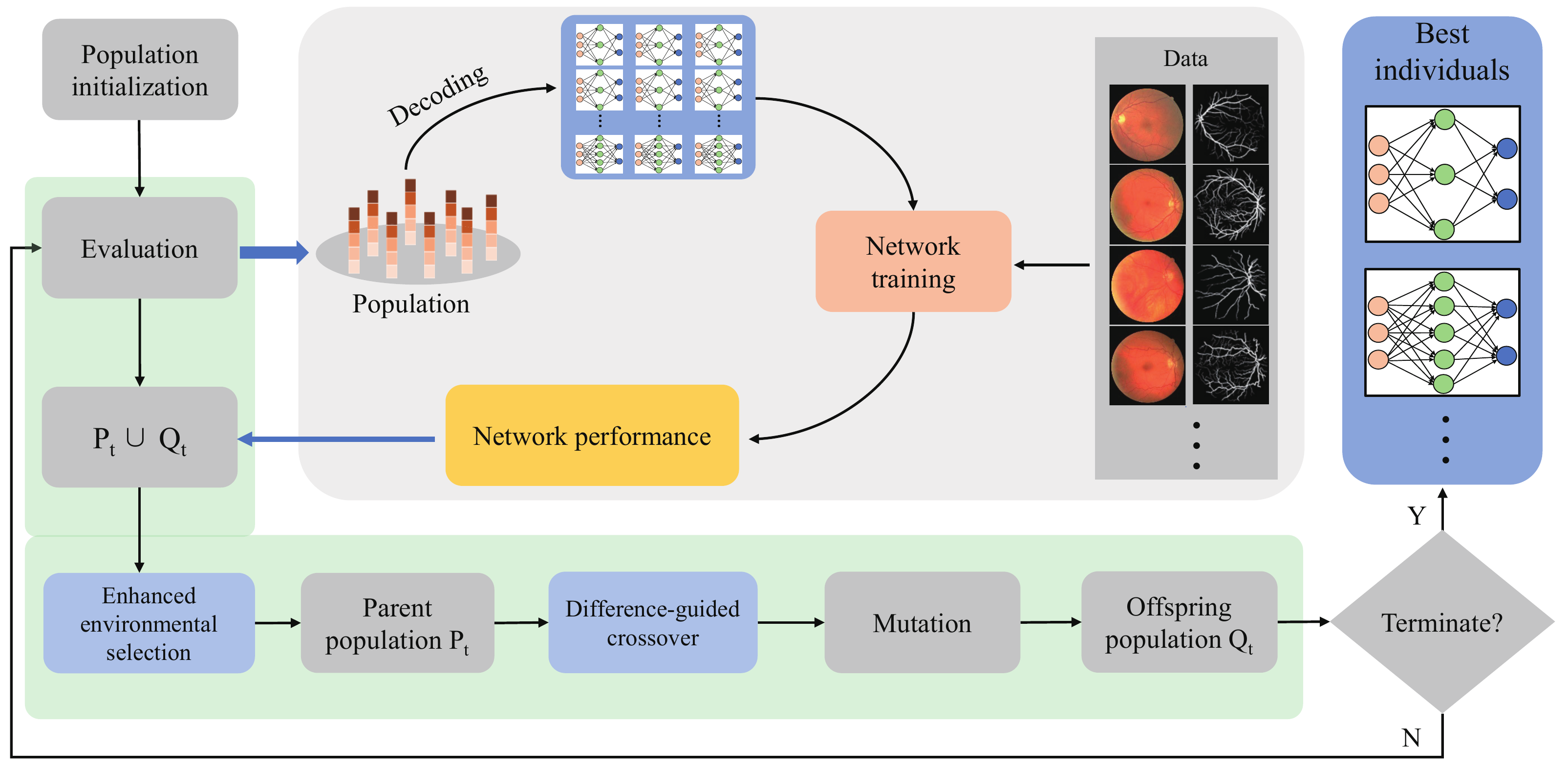}
	\caption{The overall framework of the proposed method.}\label{flowcart}
	
\end{figure*}

Genetic U-Net follows an iterative evolutionary process to generate a continuously improved population. In the population, each individual stands for architecture, and its fitness depends on the performance of the corresponding architecture in particular applications.
The Genetic U-Net flowchart is summarized in Fig. \ref{flowcart}, and the pseudocode of the proposed method is detailed in Algorithm \ref{flowcart}.
It starts with a randomly initialized population of $N$ individuals. After initialization, we evolve $T$ generations, each of which contains three evolutionary operations (e.g., crossover, mutation, and selection). Once the new individuals are generated, we evaluate them by training the architectures they encode from scratch on the provided dataset.

\normalem
\begin{algorithm}
	\KwIn{
		The population size $N$, the maximal generation number $T$, the crossover probability $p_c$, the mutation probability $p_m$, the mutation probability $p_b$ of each bit.
	}
	\KwOut{The discovered best architectures.}
	$P_0 \leftarrow$ Randomly initialize a population of $N$ by \textbf{\textit{the designed encoding strategy}};\\
	Evaluate the fitness of individuals in $P_0$;\\
	\For{$t=1$ to $T$}{
		$Q_t\leftarrow\emptyset$;\\
		\While{$\left | Q_t \right |< N$}{
			$o_1$, $o_2$ $\leftarrow$ Generate two offspring by \textbf{\textit{the designed difference-guided crossover}} operation with the probability $p_c$ from $P_{t-1}$;\\
			$o_1$, $o_2$ $\leftarrow$ Apply the mutation with the mutation probability $p_m$ and a flipping probability of $p_b$ to $o_1$ and $o_2$;\\
			$Q_t\leftarrow Q_t\cup o_1\cup o_2$;\\
		}
		Evaluate the fitness of individuals in $Q_t$;\\
		$P_{t}\leftarrow$ Select $N$ individuals from $P_{t-1} \cup Q_t$ using \textbf{\textit{the proposed environmental selection}};\\
	}
	\Return the individuals with the best fitness in $P_t$.
	\caption{Framework of the Proposed Method}
	\label{alg:1}
\end{algorithm}

\subsubsection{Crossover Operation}
\label{sec:3.2.1}
Crossover is intended to exchange information between individuals, and effective information exchange can improve the performance of the algorithm.
Many studies \cite{de1992formal,lin1997analysing} have shown the benefits of crossover operators involving a higher number of crossover points; consequently, a multi-point crossover is adopted herein.
To improve search efficiency, we design a method named \textit{\textbf{difference-guided}} crossover to choose two relatively different parent individuals for crossover. Algorithm \ref{alg:2} shows the details of the crossover operation. Initially, two individuals, $p_{1}$ and $p_{2}$, are chosen by binary tournament selection \cite{miller1995genetic}. 
Next, the Hamming distance \cite{hamming1950error} between $p_{1}$ and $p_{2}$ is calculated and normalized to $\left [0,1 \right ]$ and denoted \textit{diff}.
If \textit{diff} is larger than the pre-defined threshold $\mu$, $p_{1}$ and $p_{2}$ are designated as parents. Otherwise, $p_{1}$ and $p_{2}$ will be re-selected by again using binary tournament selection. If ten re-selections of $p_{1}$ and $p_{2}$ do not meet the requirement, the last selection will be designated as parents. After that, the designated parents will mate with the probability $p_{c}$.

\begin{algorithm}
	\KwIn{
		The population $P_t$, the probability for crossover operation $p_c$, the difference threshold of crossover operation $\mu$.
	}
	\KwOut{Two offspring $o_1$, $o_2$.}
	$o_1$, $o_2 \leftarrow \emptyset $ \\
	\For{$j=1$ to $10$}{
		$p_1, p_2 \leftarrow$ Select two individuals from $P_t$ by binary tournament selection;\\
		$\textit{diff} \leftarrow$ Compute the difference between $p_1$ and $p_2$;\\
		\If{$\textit{diff} > \mu$}{
			\textbf{break};
		}
	}
	$r \leftarrow$ Randomly generate a number from (0, 1);\\
	\eIf{$r<p_c$}{
		$len \leftarrow$ Compute the length of $p_1$ and $p_2$;\\
		$ints \leftarrow$ Randomly choose ten different integers from $[0, len)$ and sort them;\\
		$(i_0, i_1)$, $(i_2, i_3)$, $(i_4, i_5)$, $(i_6, i_7)$, $(i_8, i_9) \leftarrow$ Divide $ints$ into five pairs in order;\\
		\For{$k=1$ to $5$}{
			$p_1$, $p_2 \leftarrow$ Exchange $p_{1}[i_{2k-2}: i_{2k-1}]$ and $p_{2}[i_{2k-2}: i_{2k-1}]$;\\
		}
		$o_1$, $o_2 \leftarrow$ $p_1$, $p_2$;\\
	}{$o_1$, $o_2 \leftarrow$ $p_1$, $p_2$;}
	
	\Return $o_1$, $o_2$.
	\caption{Difference-guided Crossover Operation}
	\label{alg:2}
\end{algorithm}
\subsubsection{Mutation}
\label{sec:3.2.2}
Mutation can promote population diversity and prevent the algorithm from being trapped in a local optimum. In the proposed method, the offspring generated via crossover have the probability of $p_{m}$ for mutation, and each bit has the probability of $p_b$ for flipping independently. $p_b$ is a relatively small value (e.g., 0.05) , which prevents an individual from being altered too much by mutation.

\subsubsection{Environmental Selection}
\label{sec:3.2.3}
Typically, GAs select the next population by tournament or roulette selection. Both selection methods may miss the best individual, resulting in performance degradation \cite{holland1992adaptation}, which considerably impacts the final optimization results. Conversely, if we explicitly select the top N individuals as the next generation, a premature phenomenon \cite{leung1997degree,zbigniew1996genetic} is more likely to trap the algorithm in a local optimum \cite{goldberg1988genetic} because of the heavy loss of population diversity. Hence, when choosing the next population, both elitist and non-elitist individuals are selected, thereby improving the convergence of the algorithm while maintaining the population diversity.

Algorithm \ref{alg:3} shows the process of environmental selection for the algorithm. First, given the current population $P_t$ and the generated offspring population $Q_t$, the top five best individuals are selected into the next population $P_{t+1}$ and removed from $P_t\cup Q_t$. Second, $\left | P_t \right | - 5$ individuals are selected from $P_t\cup Q_t$ using the binary tournament selection, and then placed into $P_{t+1}$. The size of the next population $P_{t+1}$ is thus kept to that of the current population $P_t$.
\begin{algorithm}
	\KwIn{
		The parent population $P_t$, the offspring population $Q_t$.
	}
	\KwOut{The population for the next generation $P_{t+1}$.}
	$P_{t+1} \leftarrow \emptyset$;\\
	$P_{best} \leftarrow$ Select the best five individuals from $P_t \cup Q_t$;\\
	$P_t \cup Q_t  \leftarrow$ Remove $P_{best}$ from $P_{t} \cup Q_{t}$;\\
	$P_{t+1}  \leftarrow P_{t+1} \cup P_{best}$;\\
	\While{$\left | P_{t+1} \right | < \left | P_t \right |$}{
		$p \leftarrow$ Select an individual from $P_t$ by binary tournament selection;\\
		$P_{t+1} \leftarrow P_{t+1} \cup p$;\\
	}
	\Return $P_{t+1}$.
	\caption{Environmental Selection}
	\label{alg:3}
\end{algorithm}

\subsubsection{Fitness Evaluation}
\label{sec:3.2.4}
In Genetic U-Net, the fitness of an individual is the \textit{F1} (as explained in section \ref{sec:4.3}) based on the architecture the individual represents and the data for validation because the \textit{F1} is a comprehensive metric for retinal vessel segmentation that helps resolve the problem of imbalanced samples. Algorithm \ref{alg:4} summarizes the procedure for evaluating individuals in the population. For evaluation, each individual transforms itself into an architecture, which is an inverse binary encoding process. Before training, He initialization \cite{he2015delving} is used to initialize the architecture’s weights. Then, on the training data, the architecture is trained by Lookahead \cite{zhang2019lookahead}, which uses Adam \cite{kingma2014adam} as the base optimizer. After 80 training epochs, the validation data are used to validate the trained architecture after each epoch until the $130^{th}$ epoch, and the best \textit{F1} during this process is set as the fitness of the corresponding individual.
\begin{algorithm}
	\KwIn{
		The population $P_t$ for fitness evaluation, training data $D_{train}$, validation data $D_{valid}$.
	}
	\KwOut{The population $P_t$ with fitness.}
	\ForEach{ $individual$ in $P_t $}{
		$arch \leftarrow$ Transform the $individual$ to its corresponding architecture;\\
		Apply weight initialization to $arch$;\\
		$F1_{best} \leftarrow 0$;\\
		\For{$epoch=1$ to $130$}{
			Train $arch$ on $D_{train}$ by gradient descent for an epoch;\\	
			\If{$epoch > 80$}{
				$F1 \leftarrow$ Evaluate the trained $arch$ on $D_{valid}$;\\
				\If{$F1>F1_{best}$}{
					$F1_{best} \leftarrow F1$;\\	
				}
			}
		}
		Set $F1_{best}$ as the fitness of $individual$;\\
	}
	\Return $P_{t}$.
	\caption{Evaluate Fitness}
	\label{alg:4}
\end{algorithm}

\section{Materials for Experiments}
\label{sec:4}
\subsection{Loss Function}
\label{sec:4.1}
In fundus images, the ratio of vessel pixels is less than 0.1, and most pixels belong to the non-vessel class. Therefore, the problem of imbalanced samples must be resolved. For this purpose, focal loss \cite{lin2017focal} originally proposed to alleviate the sample imbalance problem in object detection is utilized as the loss function in this work, which is given as 
\begin{equation}\label{eq:loss}
\begin{split}
Loss=-\sum_{n=1}^{m} ( \alpha y_n \left ( 1-\hat{y}_n \right )^{\omega }log\hat{y}_n \\
+\left ( 1-\alpha   \right )\left ( 1-y_n \right )\hat{y}_n^\omega log\left (1-\hat{y}_n  \right ) )
\end{split}
\end{equation}
where $y$, $\hat{y}$, $n$, $m$, $\alpha$, and $\omega$ respectively indicate the ground truth, model prediction, $n^{th}$ sample, sample total, balance factor between positive and negative samples, and balance factor between simple and hard samples.

\subsection{Datasets}
\label{sec:4.2}
Four public datasets DRIVE \cite{staal2004ridge}, CHASE\_DB1 \cite{owen2009measuring}, STARE \cite{hoover2000locating}, and HRF \cite{kohler2013automatic} were used in our experiments. Descriptions of these datasets are given in Table \ref{tab:dataset}. The images of these datasets were captured by different devices and with different image sizes. DRIVE, STARE, and CHASE\_DB1 have different annotations from two experts but only the annotations of the first expert were taken as the experimental ground truth. HRF has annotations from only one expert. The training and test datasets were split using the same method as in \cite{alom2019recurrent,li2015cross,yan2018joint}. DRIVE has an officially determined training-testing split, 20 images for training and 20 for testing. For STARE, leave-one-out cross-validation was used to divide the dataset. For CHASE\_DB1, the first 20 images were used for training, and the remaining 8 images were assigned to the test set. The 45 images in HRF are divided into three categories; healthy, diabetic retinopathy, and glaucomatous, each of which contains 15 images. The first five images of every category were used for training and the remainder were used for testing. Binary field of view (FOV) masks are offered for the DRIVE and HRF datasets. FOV masks for the STARE and CHASE\_DB1 datasets were generated following \cite{liskowski2016segmenting} because they were not provided. 

\begin{table}[htbp]
	\centering
	\caption{Descriptions of the datasets.}
	\scalebox{0.7}[0.7]{\begin{tabular}{ccccc}
		\toprule[1.5pt]
		Dataset  &Quantity& Resolution & training-testing split\\
		\midrule[1.5pt]
		DRIVE  & 40 & $565\times 584$ & 20/20 \\
		STARE  & 20 & $700\times 605$ & leave one out \\
		CHASE\_DB1  & 28 & $999\times 960$ & 20/8 \\
		HRF  & 45 & $3504\times 2336$ & 15/30 \\
		\bottomrule[1.5pt]
	\end{tabular}}%
	\label{tab:dataset}%
\end{table}%

\subsection{Evaluation Metrics}
\label{sec:4.3}
Retinal vessel segmentation is a binary classification problem that classifies each pixel in the fundus image as either a vessel or a non-vessel. The output of the model is a probability map, which to each pixel assigns the probability of belonging to the class of vessels. In the experiments, the probability threshold was set to 0.5 to obtain the results. If a vessel pixel is correctly classified, it is a true-positive (\textit{TP}); if not, it is a false-positive (\textit{FP}). If a non-vessel pixel is precisely classified, it is a true-negative (\textit{TN}); if not, it is a false-negative (\textit{FN}). As shown in Table \ref{tab:metrics}, five metrics were selected for evaluation. Unless otherwise specified, we only used pixels inside FOVs to calculate the performance metrics.

\begin{table}[htbp]
	\centering
	\caption{Metrics for evaluation in our work.}
	\scalebox{0.9}[0.9]{\begin{tabular}{ll}
		\toprule[1.5pt]
		Metrics & Description \\
		\midrule[1.5pt]
		$ACC$ (\textit{accuracy}) & $ACC = (TP + TN) $ / $(TP + TN + FP + FN) $\\
		$SE$ (\textit{sensitivity}) & $SE = TP$ / $(TP + FN) $\\
		$SP$ (\textit{specificity}) & $SP = TN$ / $ (TN + FP) $\\
		\textit{F1} (\textit{F1\_score}) & \textit{F1} $= (2\times TP) $ / $(2\times TP + FP +FN) $\\
		\textit{AUROC} & Area Under the ROC Curve. \\
		\bottomrule[1.5pt]
	\end{tabular}}
	\label{tab:metrics}
\end{table}%

\section{Experiments}
\label{sec:5}
The experiments in the study consisted of two stages. 
In the first stage, the neural architectures were sought by GA.
In the second stage, the found architectures were trained from scratch (validated) to establish their performances on retinal vessel segmentation. In this section, the stages of the experiments are explained and their results analyzed.
\subsection{Experimental Setup}
\label{sec:5.1}
\textbf{Dataset for searching:} {We sought the architectures using DRIVE. In the architecture search stage (first stage), the last five images of its training set were selected for validation, while the remaining fifteen images were for training. We also tested the architectures found using DRIVE on other datasets (STARE, CHASE\_DB1, and HRF) in the second stage.}

\textbf{Genetic U-Net hyper-parameters:} 
The generic up-sampling and down-sampling operations (max pooling and transpose convolution) were adopted as for the original U-Net \cite{ronneberger2015u}.
The probability of crossover and mutation operations ($p_c$ and $p_m$) was set to 0.9 and 0.7 respectively, and the difference threshold $\mu$ was set to 0.2. The probability $P_b$ in the mutation process was 0.05. The population size $N$ was 20, and the number of the generations $T$ was 50.

\textbf{Network training during search:} For data argumentation, horizontal flipping, vertical flipping, and random rotation from $\left [ 0 \degree, 360 \degree \right ]$ were employed to increase the training data, which prevented the models from overfitting. The pixels of the images were normalized to $\left [ -0.5,0.5 \right ]$. We took the full image as the input instead of patches, and only one image was input into the model per iteration. Lookahead \cite{zhang2019lookahead} and Adam \cite{kingma2014adam} took the default parameters (e.g., $\alpha = 0.5$, $k = 6$, $\beta_1 = 0.9$, $\beta_2 = 0.999$) for optimization. The learning rate was initialized as 0.001. The architectures were trained on two NVIDIA TITAN RTX GPUs implementing PyTorch 1.5.0.

\textbf{Network training after search:} In the second stage, the training settings were similar to those used in the architecture search stage (e.g., optimizer, loss function, and data argumentation). The main difference was that the number of training epochs was expanded to 900 to ensure the convergence of the training. The data split was as described in section \ref{sec:4.2}. In addition, due to the GPU memory limitation, during training, the images in HRF were randomly cropped out of a region 1000 $\times$ 1000 in size and used as the architecture input but the complete images were adopted as the input during testing.

\subsection{Experimental Results of the Discovered Model}
\label{sec:5.2}
\subsubsection{Comparison with Existing Methods}
\label{sec:5.2.1}
We report the test results of the architectures searched for on DRIVE on all four public datasets (DRIVE, STARE, HRF, and CHASE\_DB1) and compare them with other existing methods (mostly are CNN-based methods).

The experimental results are summarized in Tables \ref{tab:DRIVE}, \ref{tab:STARE}, \ref{tab:CHASEB_DB1}, and \ref{tab:HRF}. In these tables, the results of existing methods except CE-Net \cite{gu2019net} and CS$^2$-Net \cite{mou2021cs2} are obtained from the original papers. The results of CE-Net and CS$^2$-Net are obtained using the same training settings as the discovered architecture. The existing methods marked with stars calculate the performance metrics with FOVs, while those without stars were not explicitly claimed in the papers. Here, we report the performance of the discovered architecture both with FOVs and without FOVs. For fairness, our method used the same data split as the other methods.
For all the datasets, the proposed method dominates all the methods of comparison on two comprehensive metrics (F1 and AUROC), which means that the discovered architecture achieves the best overall performance without post-processing or special pre-processing.
Additionally, the proposed method outperformed the other methods in SE (sensitivity) by some margin, which indicates that the discovered architecture is more capable of detecting vessels.
ACC is a trade-off between SE and SP. Since the non-vessel region is much larger than the vessel region in fundus images, ACC is more susceptible to SP.
The ACC of the proposed method also compares well against the other methods.
Moreover, the parameters of the discovered architecture are the least among these methods (0.27 M).
We utilized only a few data augmentations such as flipping and rotation that are simpler and less diverse than those used by most other methods. The architecture searched for on DRIVE also successfully transferred to the other three datasets, STARE, HRF, and CHASE\_DB1.

\begin{table}[htbp]
	\centering
	\caption{Comparison with existing methods on DRIVE dataset.}
	\begin{threeparttable}
	\scalebox{0.6}[0.6]{\begin{tabular}{llllllll}
		\toprule[1.5pt]
		Methods & Year  & ACC   & SE    & SP    & Fl    & AUROC  & Params(M) \\
		\midrule[1.5pt]
		Vega \textit{et al}. \cite{vega2015retinal} & 2015  & 0.9412 & 0.7444 & 0.9612 & 0.6884 & N/A & N/A\\
		Li \textit{et al}.$^*$ \cite{li2015cross}  & 2015  & 0.9527 & 0.7569 & 0.9816 & N/A   & 0.9738 & N/A\\
		Orlando \textit{et al}.$^*$ \cite{orlando2016discriminatively} & 2016  & N/A   & 0.7897 & 0.9684 & 0.7857 & N/A & N/A\\
		Fan and Mo \cite{fan2016automated_2} & 2016  & 0.9612 & 0.7814 & 0.9788 & N/A   & N/A & N/A\\
		Liskowski \textit{et al}.$^*$ \cite{liskowski2016segmenting} & 2016  & 0.9535 & 0.7811 & 0.9807 & N/A   & 0.9790 & 48.00 \\

		Mo and Zhang \cite{mo2017multi} & 2017  & 0.9521 & 0.7779 & 0.9780 & N/A   & 0.9782 & 7.63\\
		Yan \textit{et al}.$^*$ \cite{yan2018joint}        & 2018 & 0.9542 & 0.7653 & 0.9818 & N/A    & 0.9752 & 31.35   \\
		Alom \textit{et al}.$^*$ \cite{alom2019recurrent}  & 2019  & 0.9556 & 0.7792 & 0.9813 & 0.8171 & 0.9784 & 1.07\\
		Jin \textit{et al}.$^*$ \cite{jin2019dunet} & 2019  & 0.9566 & 0.7963 & 0.9800 & 0.8237 & 0.9802 & 0.88\\
		Bo Wang \textit{et al}.$^*$ \cite{wang2019dual} & 2019  & 0.9567 & 0.7940 & 0.9816 & 0.8270 & 0.9772 & N/A\\
		Yicheng Wu \textit{et al}. \cite{wu2019vessel} & 2019  & 0.9578 & 0.8038 & 0.9802 & N/A   & 0.9821 & 1.70\\
		Mou Lei \textit{et al}. \cite{mou2019dense} & 2019  & 0.9594  & 0.8126  & 0.9788  & N/A   & 0.9796  & 56.03\\
		CE-Net$^*$ \cite{gu2019net} & 2019  & 0.9545  & 0.8276  & 0.9735  & 0.8243   & 0.9794  & 15.28\\
		CS$^2$-Net$^*$ \cite{mou2021cs2} & 2021  & 0.9553  & 0.8154  & 0.9757  & 0.8228   & 0.9784  & 8.91\\
		\midrule
		Genetic U-Net w/ FOVs  & -  & 0.9577 & \textbf{0.8300} & 0.9758 & \textbf{0.8314} & \textbf{0.9823} & \textbf{0.27} \\
		Genetic U-Net w/o FOVs  & -  & \textbf{0.9707} & \textbf{0.8300} & \textbf{0.9843} & \textbf{0.8314} & \textbf{0.9885} & \textbf{0.27} \\
		\bottomrule[1.5pt]
	\end{tabular}%
	}
\begin{tablenotes}
	\scriptsize
	\item 'W/' and 'W/o' mean 'with' and 'without', respectively.
\end{tablenotes} 
\end{threeparttable}
	\label{tab:DRIVE}%
\end{table}%

\begin{table}[htbp]
	\centering
	\caption{Comparison with existing methods on the STARE dataset.}
	\scalebox{0.7}[0.7]{\begin{tabular}{lllllll}
		\toprule[1.5pt]
		Methods            & Year & ACC    & SE     & SP     & Fl     & AUROC  \\
		\midrule[1.5pt]
		Vega \textit{et al}. \cite{vega2015retinal}       & 2015 & 0.9483 & 0.7019 & 0.9671 & 0.6614 & N/A    \\
		Li \textit{et al}.$^*$ \cite{li2015cross}       & 2015 & 0.9628 & 0.7726 & 0.9844 & N/A    & 0.9879 \\
		Orlando \textit{et al}.$^*$ \cite{orlando2016discriminatively}  & 2017 & N/A    & 0.7680 & 0.9738 & 0.7644 & N/A    \\
		Fan and Mo \cite{fan2016automated_2}       & 2016 & 0.9654 & 0.7834 & 0.9799 & N/A    & N/A    \\
		Liskowski \textit{et al}.$^*$ \cite{liskowski2016segmenting} & 2016 & 0.9566 & 0.7867 & 0.9754 & N/A    & 0.9785 \\
		Mo and Zhang \cite{mo2017multi}    & 2018 & 0.9674 & 0.8147 & 0.9844 & N/A    & 0.9885 \\
		Yan \textit{et al}.$^*$ \cite{yan2018joint}        & 2018 & 0.9612 & 0.7581 & 0.9846 & N/A    & 0.9801    \\
		Alom \textit{et al}.$^*$ \cite{alom2019recurrent}     & 2019 & 0.9712 & 0.8292 & 0.9862 & 0.8475 & 0.9914 \\
		Jin \textit{et al}.$^*$ \cite{jin2019dunet}        & 2019 & 0.9641 & 0.7595 & 0.9878 & 0.8143 & 0.9832 \\
		CE-Net$^*$ \cite{gu2019net} & 2019  & 0.9656  & 0.8406  & 0.9813  & 0.8363   & 0.9871  \\
		CS$^2$-Net$^*$ \cite{mou2021cs2} & 2021  & 0.9670  & 0.8396  & 0.9813  & 0.8420   & 0.9875  \\
		\midrule
		Genetic U-Net w/ FOVs  & - & \textbf{0.9719} & \textbf{0.8658} & 0.9846 & \textbf{0.8630} & \textbf{0.9921}\\
		Genetic U-Net w/o FOVs  & - & \textbf{0.9792} & \textbf{0.8658} & \textbf{0.9886} & \textbf{0.8630} & \textbf{0.9942}\\
		\bottomrule[1.5pt]
	\end{tabular}}
	\label{tab:STARE}%
\end{table}

\begin{table}[htbp]
	\centering
	\caption{Comparison with existing methods on the CHASE\_DB1 dataset.}
	\scalebox{0.65}[0.65]{\begin{tabular}{lllllll}
		\toprule[1.5pt]
		Methods & Year  & ACC   & SE    & SP    & Fl    & AUROC \\
		\midrule[1.5pt]
		Li \textit{et al}.$^*$ \cite{li2015cross} & 2015  & 0.9581 & 0.7507 & 0.9793 & N/A   & 0.9716 \\
		Fan and Mo \cite{fan2016automated_2} & 2016  & 0.9573 & 0.7656 & 0.9704 & N/A   & N/A \\
		Yan \textit{et al}.$^*$ \cite{yan2018joint} & 2018  & 0.9610 & 0.7633 & 0.9809 & N/A   & 0.9781 \\
		Alom \textit{et al}.$^*$ \cite{alom2019recurrent} & 2019  & 0.9634 & 0.7756 & 0.9820 & 0.7928 & 0.9815 \\
		Bo Wang \textit{et al}.$^*$ \cite{wang2019dual} & 2019  & 0.9661 & 0.8074 & 0.9821 & 0.8037 & 0.9812 \\
		Yicheng Wu \textit{et al}. \cite{wu2019vessel} & 2019  & 0.9661 & 0.8132 & 0.9814 & N/A   & 0.9860  \\
		CE-Net$^*$ \cite{gu2019net} & 2019  & 0.9641  & 0.8093  & 0.9797  & 0.8054   & 0.9834  \\
		CS$^2$-Net$^*$ \cite{mou2021cs2} & 2021  & 0.9651  & 0.8329  & 0.9784  & 0.8141  & 0.9851  \\
		\midrule
		Genetic U-Net w/ FOVs  & -  & \textbf{0.9667 } & \textbf{0.8463 } & 0.9845 & \textbf{0.8223} & \textbf{0.9880 } \\
		Genetic U-Net w/o FOVs  & -  & \textbf{0.9769 } & \textbf{0.8463 } & \textbf{0.9857 } & \textbf{0.8223 } & \textbf{0.9914 } \\
		\bottomrule[1.5pt]
	\end{tabular}}%
	\label{tab:CHASEB_DB1}%
\end{table}%

\begin{table}[htbp]
	\centering
	\caption{Comparison with existing methods on the HRF dataset.}
	\scalebox{0.7}[0.7]{\begin{tabular}{lllllll}
			\toprule[1.5pt]
			Methods & Year  & ACC   & SE    & SP    & Fl    & AUROC \\
			\midrule[1.5pt]
			Orlando \textit{et al}.$^*$ \cite{orlando2016discriminatively} & 2016  & N/A   & 0.7794 & 0.9650 & 0.7341 & N/A \\
			Yan \textit{et al}.$^*$ \cite{yan2018joint} & 2018  & 0.9437 & 0.7881 & 0.9592 & N/A   & N/A \\
			Jin \textit{et al}.$^*$ \cite{jin2019dunet} & 2019  & 0.9651 & 0.7464 & \textbf{0.9874} & N/A & 0.9831 \\
			CE-Net$^*$ \cite{gu2019net} & 2019  & 0.9613  & 0.7805  & 0.9798  & 0.7895   & 0.9766  \\
			CS$^2$-Net$^*$ \cite{mou2021cs2} & 2021  & 0.9618  & 0.7890  & 0.9795  & 0.7935   & 0.9758  \\
			\midrule
			Genetic U-Net w/ FOVs  & -  & \textbf{0.9667} & \textbf{0.8220} & 0.9818  & \textbf{0.8179 } & \textbf{0.9872} \\
			Genetic U-Net w/o FOVs & -  & \textbf{0.9715} & \textbf{0.8220} & 0.9839  & \textbf{0.8179 } & \textbf{0.9891} \\
			\bottomrule[1.5pt]
	\end{tabular}}%
	\label{tab:HRF}%
\end{table}%

\subsubsection{Comparison with the Baseline Models}
\label{sec:5.2.2}

As this work uses the U-shaped encoder-decoder structure as the backbone, we comprehensively compared the discovered architecture with the baseline models, the original U-Net \cite{ronneberger2015u}, and FC-Densenet \cite{jegou2017one}, which is also a U-shaped network.
For a fair comparison, we trained the original U-Net and FC-Densenet with the same parameter settings as the discovered architecture. Table \ref{tab:com_unet} reveals that the discovered architecture outperformed the original U-Net and FC-Densenet on all four datasets.

\begin{table}[htbp]
	\centering
	\caption{Comparison whit U-Net and FC-Densenet on four datasets.}
	\scalebox{0.66}[0.66]{\begin{tabular}{lllllll}
		\toprule[1.5pt]
		Datasets & Models & ACC   & SE    & SP    & F1    & AUROC \\
		\midrule[1.5pt]
		\multirow{2}[4]{*}{DRIVE} & U-Net & 0.9550  & 0.8091  & 0.9720 & 0.8191  & 0.9795  \\
		\cmidrule{2-7}          & FC-Densenet & 0.9555 & 0.8231 & 0.9743  & 0.8235 & 0.9799 \\
		\cmidrule{2-7}          & Genetic U-Net & \textbf{0.9577 } & \textbf{0.8300 } & \textbf{0.9758}  & \textbf{0.8314 } & \textbf{0.9823 } \\
		\midrule
		\midrule
		\multirow{2}[4]{*}{STARE} & U-Net & 0.9654 & 0.8187  & 0.9798 & 0.8273 & 0.9862 \\
		\cmidrule{2-7}          & FC-Densenet & 0.9673 & 0.8276 & 0.9805  & 0.8369 & 0.9872 \\
		\cmidrule{2-7}          & Genetic U-Net & \textbf{0.9719} & \textbf{0.8658} & \textbf{0.9846} & \textbf{0.8630 } & \textbf{0.9921} \\
		\midrule
		\midrule
		\multirow{2}[4]{*}{CHASE\_DB1} & U-Net & 0.9650  & 0.8298  & 0.9829  & 0.8092  & 0.9859  \\
		\cmidrule{2-7}          & FC-Densenet & 0.9654 & 0.8307 & 0.9789  & 0.8144 & 0.9862 \\
		\cmidrule{2-7}          & Genetic U-Net & \textbf{0.9667 } & \textbf{0.8463 } & \textbf{0.9818}  & \textbf{0.8223 } & \textbf{0.9880 } \\
			\midrule
		\midrule
		\multirow{2}[4]{*}{HRF} & U-Net & 0.9647  & 0.8212  & 0.9794  & 0.8101  & 0.9849  \\
		\cmidrule{2-7}          & FC-Densenet & 0.9652 & 0.8098 & 0.9811  & 0.8125 & 0.9851 \\
		\cmidrule{2-7}          & Genetic U-Net & \textbf{0.9667} & \textbf{0.8220} & \textbf{0.9818}  & \textbf{0.8179} & \textbf{0.9872} \\
		\bottomrule[1.5pt]
	\end{tabular}}%
	\label{tab:com_unet}%
\end{table}%

\begin{figure*}[htbp]
	\centering
	\includegraphics[width=18cm]{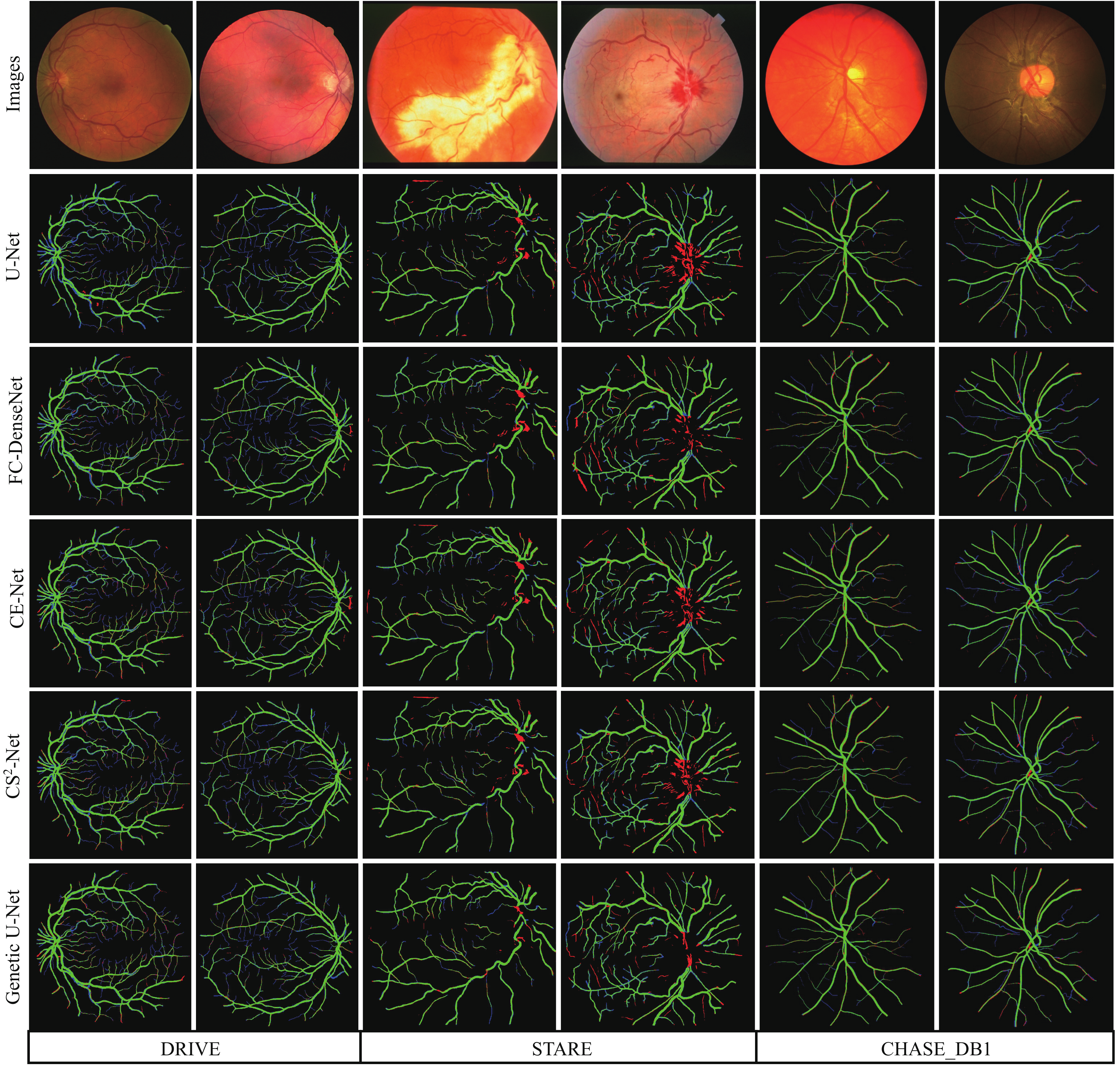}
	\caption{Overall view visualization of the segmentation results. The green pixels indicate true positive, red pixels indicate false positive, and blue pixels indicate false negative.}\label{overall-view}
	
\end{figure*}

\begin{figure}[htbp]
	\centering
	\includegraphics[width=8.5cm]{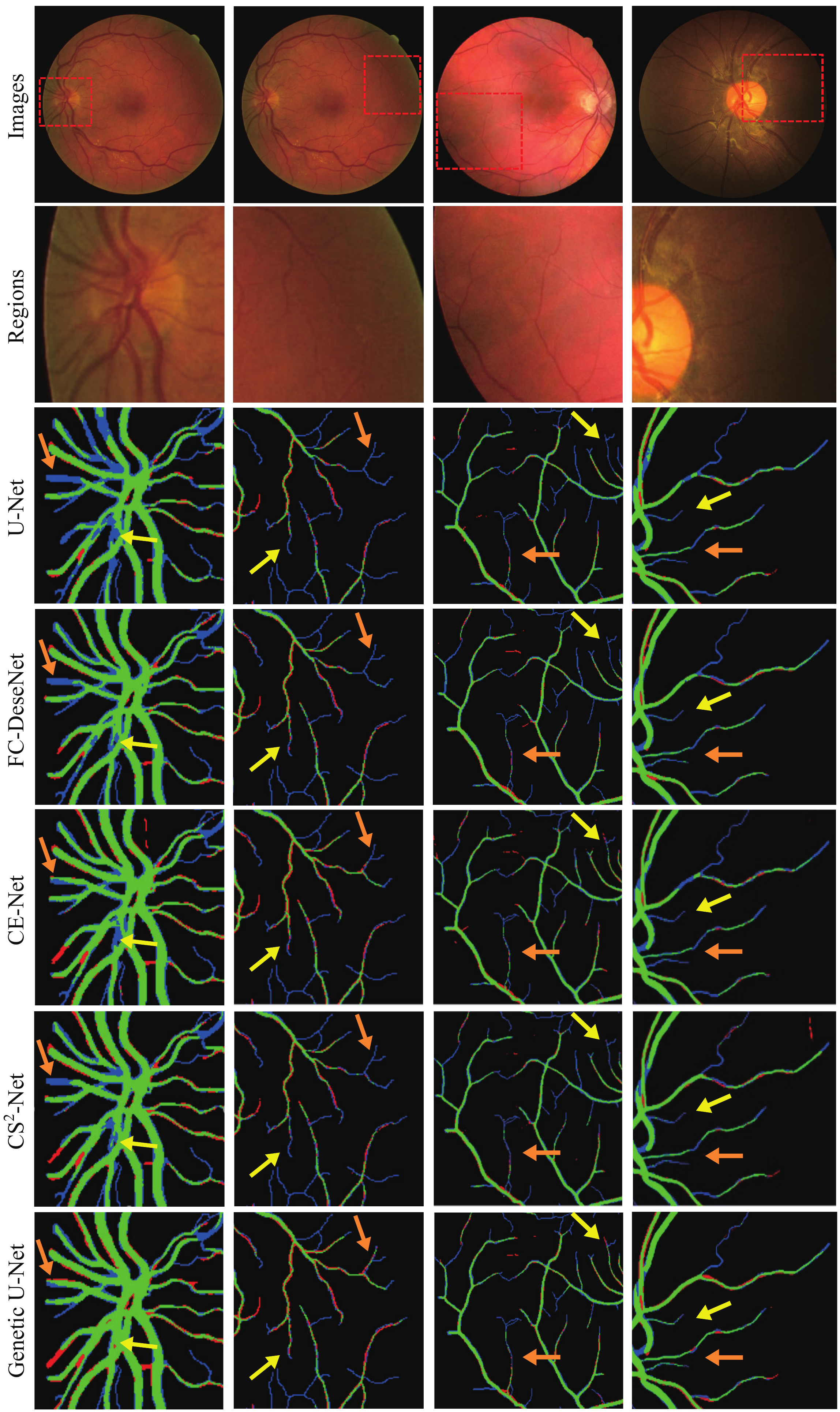}
	\caption{Locally magnified view visualization of the segmentation results.}\label{zoom-in}
	
\end{figure}

Some examples of the results are also presented in Fig. \ref{overall-view}, Fig. \ref{zoom-in}, and Fig. \ref{fig:hrf}. The blue pixels in the images indicate false negatives, which are from the vessel regions not detected. It shows that there are more blue pixels in the results of the original U-Net and FC-Densenet, in both the overall view and the locally magnified view. It further shows that the original U-Net and FC-Densenet have limitations in extracting complicated structural features, while the discovered architecture extracts them more effectively.

The computational efficiency of the discovered architecture is also analyzed, and the results are provided in Table \ref{tab:computation}. Based on Table \ref{tab:computation}, the parameters in the discovered architecture total approximately 0.27 Million, which is a 115$\times$/34$\times$ reduction compared to the 31.03 Million/9.32 Million parameters for U-Net/FC-Densenet. The model size of the discovered architecture is 1.2 MB, which is approximately a 100$\times$/32$\times$ reduction compared to the 120MB/38MB of U-Net/FC-Densenet. The execution time and Multiplication Accumulation operations (MACs) results on DRIVE were obtained with an input dimension 3$\times$565$\times$584. The total number of MACs in the discovered model in one forward propagation is 41 Billion, which is a 3.2$\times$/2.0$\times$ reduction compared to the 142 Billion/83 Billion MACs in U-Net/FC-Densenet. For the execution time, the discovered architecture also achieved an approximately 1.3$\times$/2.9$\times$ reduction compared to U-Net/FC-Densenet.

\begin{table}[htbp]
	\centering
	\caption{Comparison of model size, parameters, MACs, and execution time for U-Net and FC-Densenet. The MACs and execution time are calculated based on an input size of 3$\times$565$\times$584.}
	\scalebox{0.8}[0.8]{\begin{tabular}{lllll}
			\toprule[1.5pt]
			Models & Model size & Params & MACs & Execution time \\
			\midrule[1.5pt]
			U-Net & 120 MB & 31.03 M & 142 B & 35.4 ms \\
			FC-Densenet & 38 MB & 9.32 M & 83 B & 80.7 ms \\
			Genetic U-Net & \textbf{1.2 MB} & \textbf{0.27 M} & \textbf{41 B}  & \textbf{27.5 ms} \\
			\bottomrule[1.5pt]
	\end{tabular}}%
	\label{tab:computation}%
\end{table}%

\subsection{Experimental Analysis of Architecture Search}
\label{sec:5.3}
\begin{figure*}[htbp]
	\centering
	\includegraphics[width=18cm]{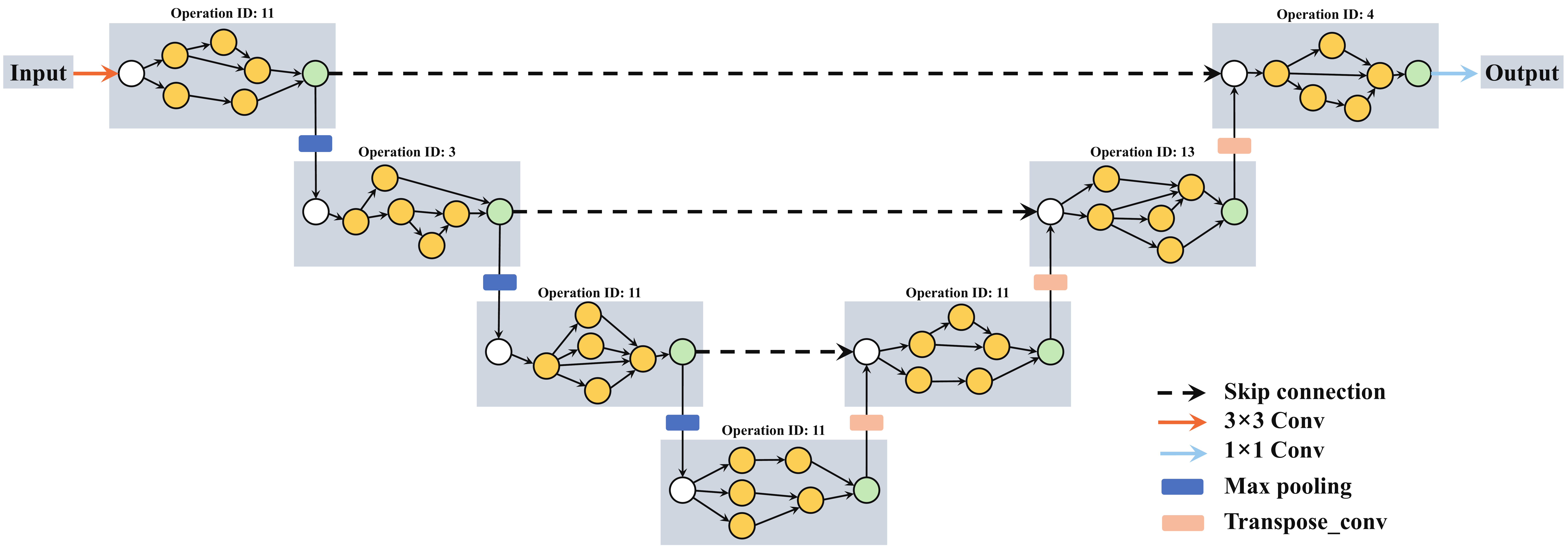}
	\caption{The first best architecture.}\label{fig:top1-arch}
	
\end{figure*}

The evolutionary trajectory of Genetic U-Net is the blue line shown in Fig. \ref{fig:trajectories}., The fitness of the best individual gradually increases from the first generation and converges at approximately the $50^{th}$ generation. Thus, we terminate the architecture search after the $50^{th}$ generation and select the best individual from the last population as the result. Fig. \ref{fig:top1-arch} displays the discovered architecture decoded from the selected individual.

\subsubsection{Observations and Findings}
\label{sec:5.3.1}
The results of the evolutionary algorithm usually contain some useful information for us to further improve our work. To find some patterns for more efficient architecture design, we observe and analyze the top five architectures of the last generation.

\textbf{Topology structure:} We observe the topology within each block of these architectures displayed in the Appendix. Almost all these blocks have the allowed maximum number of nodes, and their internal structures are relatively complex. Several skip connections can be found between the nodes. In addition, all blocks have two or three parallel branches within that resemble InceptionNet \cite{szegedy2015going} blocks.

\begin{figure}[htbp]
	\centering
	\includegraphics[width=9cm]{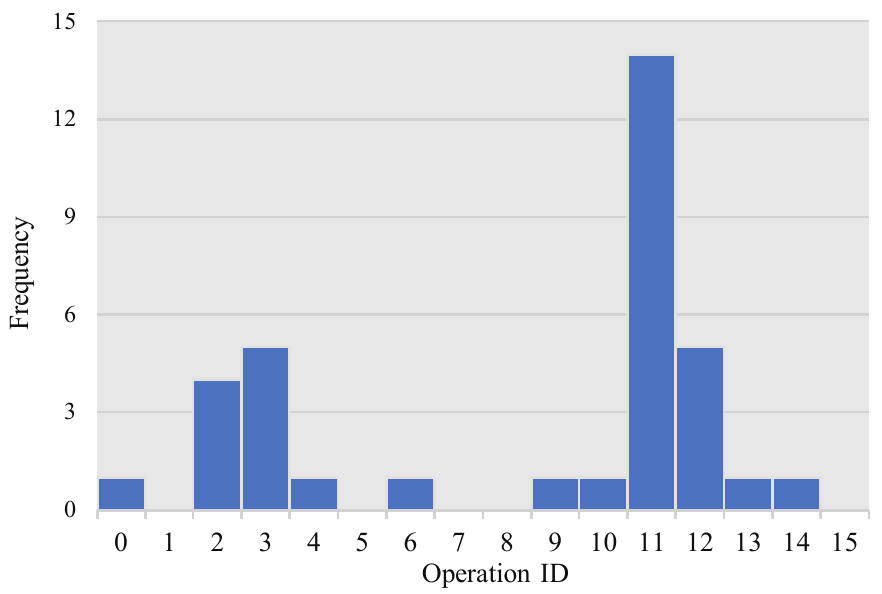}
	\caption{Frequency of the operational sequences.}\label{fig:frequency}
\end{figure}

\textbf{Operations and operational sequences:} We obtain the statistics of the frequency of occurrence of all operational sequences listed in Table \ref{tab:operations}. Fig. \ref{fig:frequency} shows that the operation sequence with ID 11 has the highest frequency. To further verify the effect of this operation sequence, we use it to replace the basic convolutional layer ($3\times3$ conv + ReLU) of the U-Net \cite{ronneberger2015u} blocks, which almost do not change parameters and MACs of U-Net, and conduct experiments on DRIVE. The result is reported in the eighth row of Table \ref{tab:exploration}. As expected, after U-Net used this operation sequence, its performance of retinal vessel segmentation on DRIVE significantly improved. Surprisingly though, its performance surpassed existing methods listed in Table \ref{tab:DRIVE}, except the discovered architecture using the proposed approach in this study. It is important to explore the reasons why only changing some operations of U-Net blocks can lead to such a significant improvement of its performance. After replacing the original basic convolution layer in U-Net with the operation sequence with ID 11, U-Net has a different activation function (Mish \cite{misra2019mish}) and activation type (pre-activation) and instance normalization, which further boosts its performance. To discover which factor or factors will have a greater impact on U-Net, we conducted six additional sets of experiments with the results listed in the second to the seventh row in Table \ref{tab:exploration}. From the data in the second to the fourth rows, adding instance normalization or using pre-activation is shown to improve the performance somewhat. In particular, the effect of adding instance normalization is apparent. U-Net is not improved by only changing the activation function to Mish. Furthermore, the data of the fifth to the eighth rows in Table \ref{tab:exploration} show that adding instance normalization or using pre-activation is useful for promoting better performance. When instance normalization and pre-activation are employed together, the effect is more notable. In addition, pre-activation with Mish is slightly more effective than pre-activation with ReLU.
\begin{table}[htbp]
	\centering
	\caption{Experimental results of verifying the operations or operational sequences.}
	\scalebox{0.8}[0.8]{
	\begin{threeparttable}
	\begin{tabular}{llclll}
		\toprule[1.5pt]
		No.   & Experiments &Params & F1 & AUROC & AUROC $^*$ \\
		\midrule[1.5pt]
		1     & U-Net-ReLU & 31.03 M& 0.8191 & 0.9795& 0.9863 \\
		2     & U-Net-Mish & 31.03 M& 0.7921 & 0.9670& 0.9801 \\
		3     & U-Net-IN-ReLU & 31.03 M& 0.8288 & 0.9814 & 0.9882\\
		4     & U-Net-ReLU(P) & 31.03 M& 0.8260 & 0.9808 & 0.9881\\
		5     & U-Net-Mish(P) & 31.03 M& 0.7920 & 0.9669& 0.9797 \\
		6     & U-Net-IN-Mish & 31.03 M& 0.8284 & 0.9812& 0.9879 \\
		7     & U-Net-IN-ReLU(P) & 31.03 M& 0.8294 & 0.9814& 0.9882 \\
		8     & U-Net-IN-Mish(P) & 31.03 M& 0.8296 & 0.9818& 0.9884 \\
		\midrule
		9     & Genetic U-Net & \textbf{0.27 M} & \textbf{0.8314}& \textbf{0.9823} & \textbf{0.9885} \\
		\bottomrule[1.5pt]
	\end{tabular}%
	\label{tab:exploration}%
\begin{tablenotes}
	\footnotesize
	\item[1] "ReLU" and "Mish" indicate the activation function used in the U-Net block and "P" represents pre-activation. "IN" means adding instance normalization.
	\item[2] "$*$" indicates that the metric is calculated without FOVs.
\end{tablenotes} 
\end{threeparttable}
}
\end{table}%

Genetic U-Net can obtain multiple competitive network architectures from these evolutionary results. We can extract knowledge and several principles of network architecture design for retinal vessel segmentation through the observations and the statistical analyses of these network architectures, which may potentially be important knowledge bases to further improve our future work.

\subsection{Ablation Study}
\label{sec:5.4}
\begin{figure}[htbp]
	\centering
	\includegraphics[width=9cm]{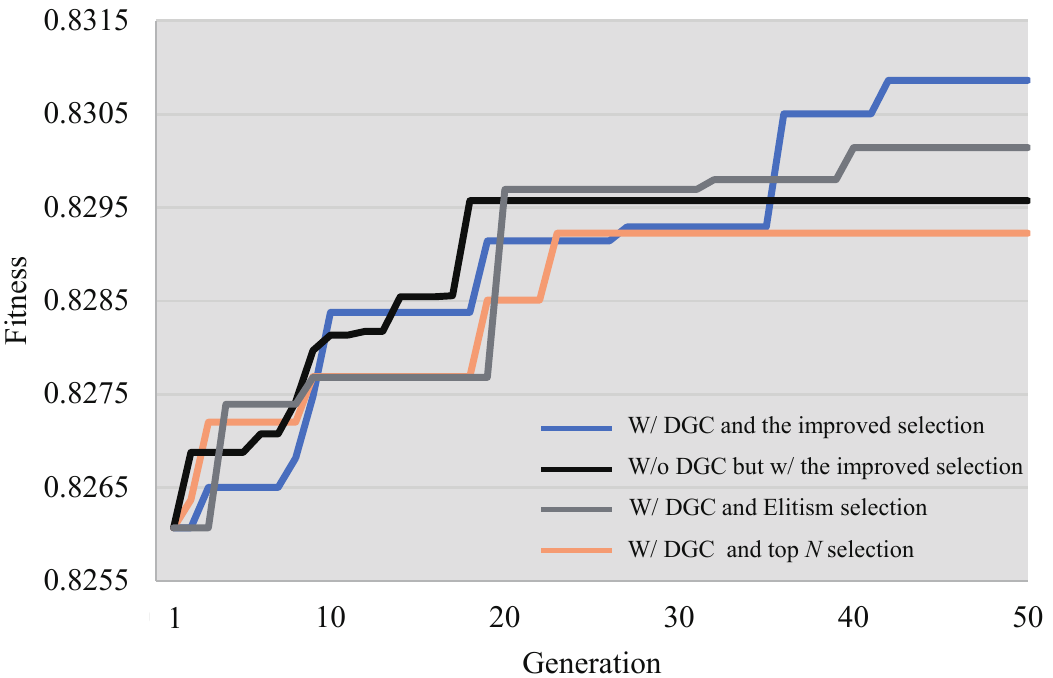}
	\caption{Evolutionary trajectories in four different situations. 'W/', 'W/o', and 'DGC' mean 'with', 'without', and 'difference-guided crossover', respectively. Top $N$ selection \cite{hancock2019selection} is to explicitly select the top $N$ best individuals for the next generation given the current population and the last population, and the population size is $N$.}\label{fig:trajectories}
	
\end{figure}

\textbf{Difference-guided crossover:} To verify the effect of utilizing the difference-guided crossover, we compare the performances of the algorithms searching with and without the difference-guided crossover. We conduct this experiment with the same settings for both algorithms. We obtain the fitness of the best individual in the population of each generation. As shown in Fig. \ref{fig:trajectories}, the difference-guided crossover improves the performance.

\textbf{Environmental selection:} To improve search efficiency and simultaneously prevent premature convergence, we adopted a selection scheme that considers both elitism (size=5) and diversity. Here, we demonstrate the benefits of this scheme by comparing it with the top $N$ selection \cite{hancock2019selection} and elitist selection (size=1) \cite{vasconcelos2001improvements}. The experiments were conducted with the same initial population and other settings (e.g., the same crossover operation.). Fig. \ref{fig:trajectories} also clearly reveals the advantage of the proposed selection scheme. The top $N$ selection results in premature convergence of the algorithm.

\section{Discussion}
\label{sec:discussion}
\subsection{Performance on Challenging Cases}
\label{sec:cases}
The images provided in the third and fourth columns of Fig. \ref{overall-view} and the second column of Fig. \ref{fig:hrf} all show lesions, and the image in the third column of Fig. \ref{overall-view} is disrupted by severely uneven illumination. In these cases, fewer red pixels (false positive or misidentified vessel regions) of lesion areas are seen in the resulting output images of the discovered architecture than those of the other models. This shows that the discovered architecture is less sensitive to lesions and uneven illumination. Similarly, as shown in the magnified view of Fig. \ref{zoom-in}, there are fewer blue pixels (false-negative or undetected vessel regions) of the cross-connected and tiny vessel areas in the resulting output images of the discovered architecture than those of the other models. This demonstrates that the discovered architecture can capture more cross-connected vessels and low-contrast tiny vessels than the experimental competitors. Overall, among the methods compared in Fig. \ref{overall-view}, Fig. \ref{fig:hrf}, and Fig. \ref{zoom-in}, the vascular trees segmented by the discovered architecture best match the ground truth. This indicates that the discovered architectures are beneficial to the segmentation of these challenging regions.

\subsection{Performance on High-Resolution Fundus Images}
\label{sec:HRF}
\begin{figure}[htbp]
	\centering
	\includegraphics[width=8.5cm]{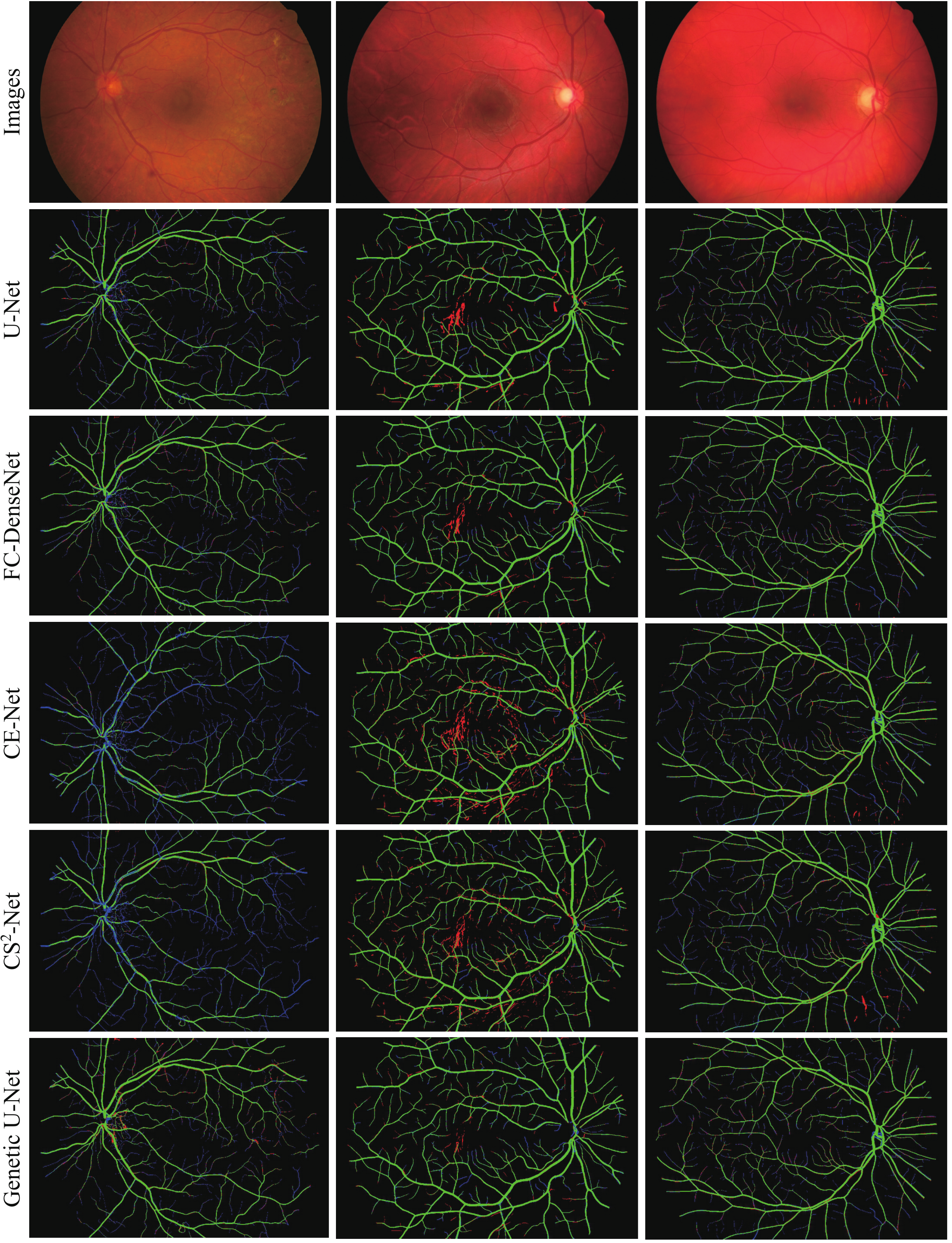}
	\caption{Segmentation results from HRF dataset. The green pixels indicate true positive, red pixels indicate false positive, and blue pixels indicate false negative.}\label{fig:hrf}
	
\end{figure}

Compared with STARE, DRIVE, and CHASE\_DB1, the HRF dataset includes fundus images with much higher resolutions. Several experimental results are displayed in Fig. \ref{fig:hrf}. The quantitative results are shown in Table \ref{tab:com_unet} and Table \ref{tab:HRF}. Both qualitative and quantitative results on HRF demonstrate that the discovered architecture achieves better segmentation results than the state-of-the-art methods and the baseline models (the original U-Net and FC-DenseNet). This indicates that the discovered architecture can also achieve promising performance on high-resolution fundus images.

\subsection{Adjusting the Number of Architecture-based Parameters}
\label{sec:achitecture parameters}
To explore the effect of model capacity further, we adjusted the channels of convolutional operations $c$ within the nodes of the discovered architecture from 20 to 2, 3, 4, 5, 10, and 30 (thus changed the number of architecture-based parameters) and trained the adjusted models from scratch. We evaluated their performance on DRIVE and CHASE\_DB1 and analyzed their computational complexity. The performance of the adjusted models is reported in Table \ref{tab:adjust}. When $c$ is increased from 20 to 30, compared with the original model, the adjusted model has more parameters and higher computational complexity (approximately a 125\% increase on MACs and Params). However, its performance was almost not improved (On DRIVE, there was an approximately 0.04\% increase in F1 and almost no increase in AUROC). When $c$ is decreased, the adjusted models have fewer parameters and lower computational complexity, and their performance does not degrade distinctly. For instance, when $c =$ 5 or 10, there is approximately a 94\% or 75\% reduction in MACs and Params, respectively. On DRIVE, however, F1 only degenerates by approximately 0.20\% or 0.91\%, and AUROC by approximately 0.13\% or 0.15\%. When $c = 5$, the adjusted model with fewer architecture-based parameters (about 0.06\%/0.19\% of the original U-Net/FC-Densenet) can still achieve competitive performance with the original U-Net, FC-Densenet, and the other methods listed in Table \ref{tab:DRIVE} and Table \ref{tab:CHASEB_DB1}.
Even if $c = 2$ and the model only contains 0.0028 Million (2.8 K) parameters, the performance is still comparable with some existing methods.
These results confirm that decreasing the capacity of the discovered architecture does not significantly influence its performance, which is beneficial for practical clinical applications. This reveals that the discovered architecture has strong parameter efficiency and is not sensitive to reductions in the number of architecture-based parameters.

\begin{table}[htbp]
\centering
\caption{Model performance after adjusting the number of architecture-based parameters.}
	\scalebox{0.6}{\begin{threeparttable} 
	\begin{tabular}{ccc|cc|cc}
	\toprule[1.5pt]
	&       &       & \multicolumn{2}{c|}{DRIVE} & \multicolumn{2}{c}{CHASE\_DB1} \\
	\midrule
	Model configs & Params & MACs  & F1 & AUROC & F1 & AUROC \\
	\midrule[1.5pt]
	c = 20 & 0.2724 M & 41.35 B & 0.8314 & 0.9823 & 0.8223 & 0.9880 \\
	c = 30  & 0.6117 M & 92.77 B & 0.8317 & 0.9824 & 0.8228 & 0.9880 \\
	c = 10  & 0.0685 M & 10.42 B & 0.8306 & 0.9815 & 0.8214 & 0.9871 \\
	c = 5   & 0.0173 M & 2.65 B & 0.8297 & 0.9810 & 0.8179 & 0.9863 \\
	c = 4   & 0.0111 M & 1.71 B & 0.8275 & 0.9806 & 0.8134 & 0.9856 \\
	c = 3   & 0.0063 M & 0.97 B & 0.8229 & 0.9791 & 0.8033 & 0.9833 \\
	c = 2   & 0.0028 M & 0.44 B & 0.8138 & 0.9746 & 0.7914 & 0.9804 \\
	\bottomrule[1.5pt]
\end{tabular}%
\label{tab:adjust}%
\begin{tablenotes}
\footnotesize
\item The MACs are calculated based on an input size of 3$\times$565$\times$584.
\end{tablenotes} 
\end{threeparttable}} 
\end{table}%

\subsection{Performance using Reduced Training Data}
\label{sec:performance using reduced training data}
\begin{figure}[htbp]
\centering
\includegraphics[width=8cm]{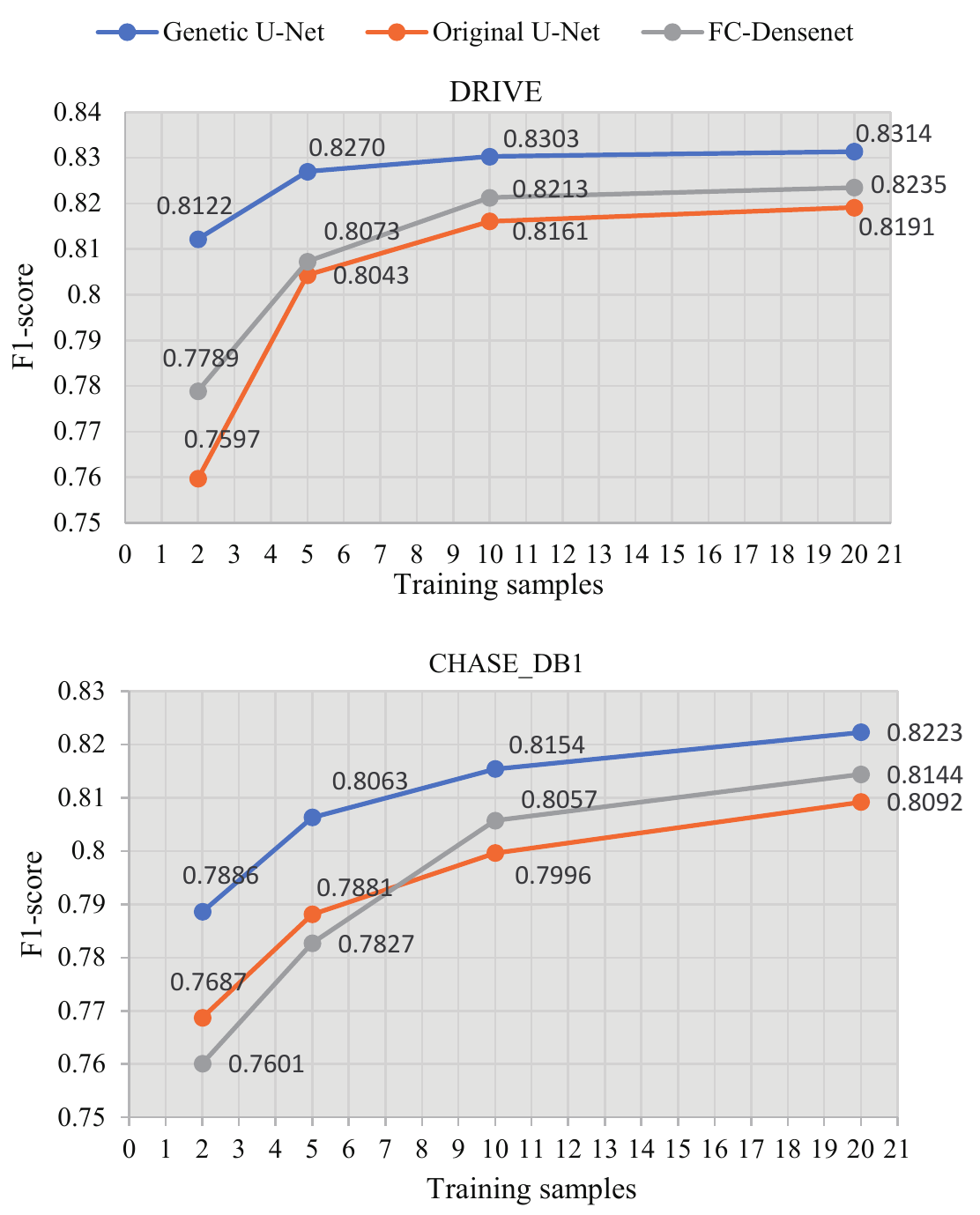}
\caption{F1 changing with the number of training images.}\label{fig:reduced}

\end{figure}

To verify the performance of the discovered architecture after reducing the training data, we evaluated it on the reduced training data of DRIVE and CHASE\_DB1. In the previous data split, both DRIVE and CHASE\_DB1 had 20 images for model training. Here, we reduced the number of images used for model training from 20 to 2, 5, and 10 in both datasets and conducted multiple experiments on the discovered architecture and the baseline models (the original U-Net and FC-Densenet). The F1 changed with the number of training images as shown in Fig. \ref{fig:reduced} and Table \ref{tab:reduced}. On DRIVE and CHASE\_DB1, both the discovered architecture and the baseline models suffered performance degradation; however, for the baseline models, the degradation is more apparent. For example, when the training data on DRIVE/CHASE\_DB1 consisted of only 5 images, the F1 of the discovered architecture decreased from 0.8314/0.8223 to 0.8270/0.8063 (0.52\%/1.94\% degradation), the F1 of the original U-Net decreased from 0.8191/0.8092 to 0.8043/0.7881 (1.80\%/2.61\% degradation), and the F1 of FC-Densenet decreased from 0.8235/0.8144 to 0.8073/0.7827 (1.97\%/3.89\% degradation).
Even with only 5 training images, the performance of the discovered architecture is still comparable with the existing methods listed in Table \ref{tab:DRIVE} and Table \ref{tab:CHASEB_DB1}. 
This confirms that the discovered architecture has more stable performance under reduced training data, which is crucial when using the limited annotated data available for retinal vessel segmentation.
\begin{table}[htbp]
	\centering
	\caption{F1 changing with the number of training images.}
	\scalebox{0.6}{\begin{tabular}{cc|cc|cc}
		\toprule[1.5pt]
		&       & \multicolumn{2}{c|}{DRIVE} & \multicolumn{2}{c}{CHASE\_DB1} \\
		\midrule
		Models & Training samples & F1 & AUROC & F1 & AUROC \\
		\midrule[1.5pt]
		\multirow{4}[2]{*}{Genetic U-Net} & 20    & 0.8314 & 0.9823 & 0.8223 & 0.9880 \\
		& 10    & 0.8303 & 0.9813 & 0.8154 & 0.9861 \\
		& 5     & 0.8270 & 0.9786 & 0.8063 & 0.9831 \\
		& 2     & 0.8122 & 0.9730 & 0.7886 & 0.9792 \\
				\midrule
		\midrule
		\multirow{4}[2]{*}{FC-Densenet} & 20    & 0.8235 & 0.9799 & 0.8144 & 0.9862 \\
		& 10    & 0.8213 & 0.9789 & 0.8057 & 0.9828 \\
		& 5     & 0.8073 & 0.9717 & 0.7827 & 0.9764 \\
		& 2     & 0.7789 & 0.9603 & 0.7601 & 0.9604 \\
		\midrule
		\midrule
		\multirow{4}[2]{*}{U-Net} & 20    & 0.8191 & 0.9795 & 0.8092 & 0.9859 \\
		& 10    & 0.8161 & 0.9754 & 0.7996 & 0.9824 \\
		& 5     & 0.8043 & 0.9700 & 0.7881 & 0.9806 \\
		& 2     & 0.7597 & 0.9514 & 0.7687 & 0.9707 \\
		\bottomrule[1.5pt]
	\end{tabular}}%
	\label{tab:reduced}%
\end{table}%

\subsection{Generalization Ability}
\label{sec:generalization ability}
\begin{figure}[htbp]
	\centering
	\includegraphics[width=8cm]{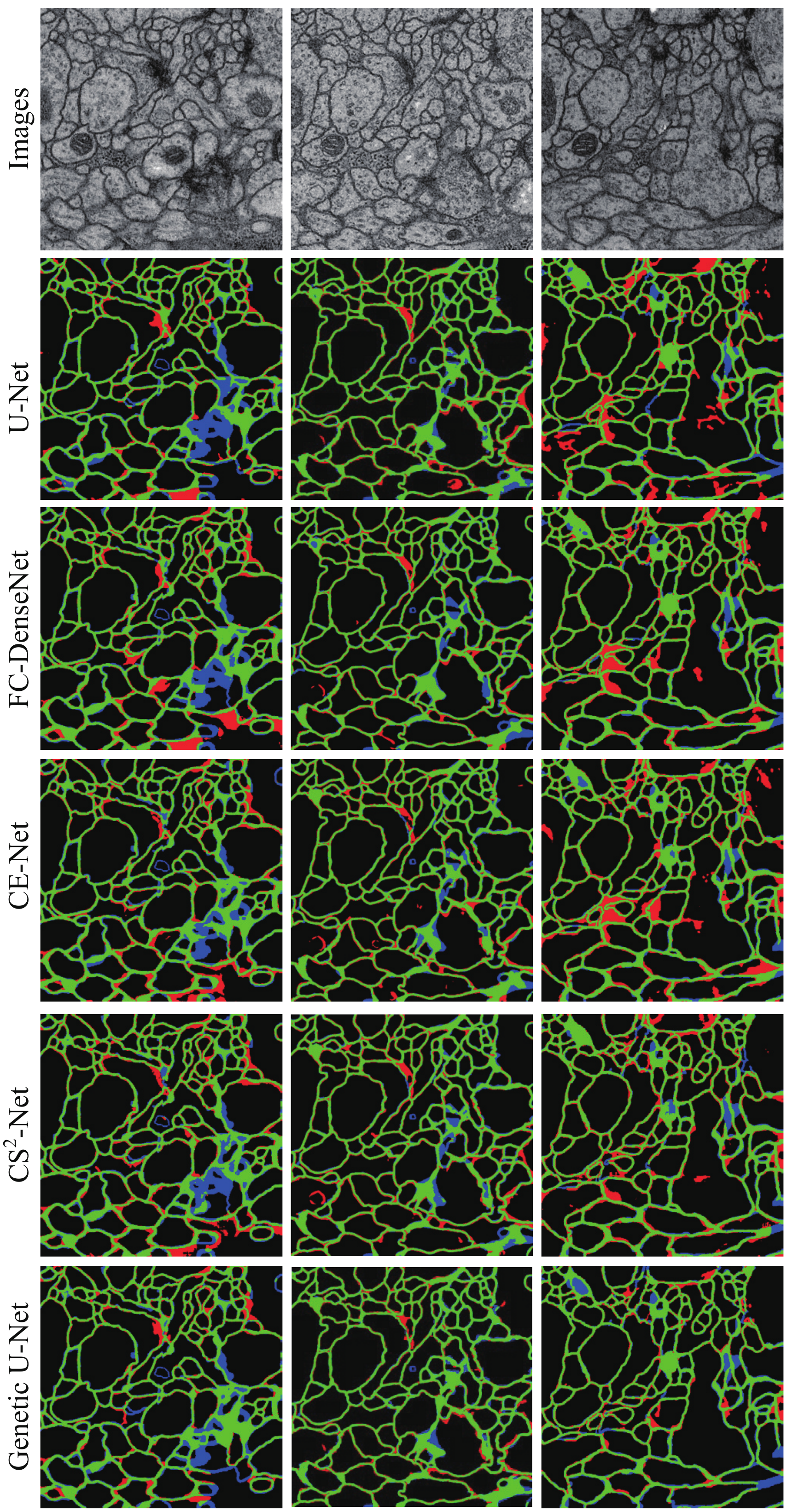}
	\caption{Visual results of cell boundary segmentation. The green pixels indicate true positive, the red pixels indicate false positive, and the blue pixels indicate false negative.}\label{fig:EM}
\end{figure}

To further evaluate the generalization ability of the discovered architecture, we tested it (searched on DRIVE) on the challenging task of cell membrane (boundary) segmentation, which is an application similar to vessel segmentation. In this experiment, the regions of the cell membranes (boundaries) are the segmented areas (positive). The adopted EM dataset \cite{cardona2010integrated} contains 30 training images and 30 test images 512 $\times$ 512 in size, but the ground truth of the 30 test images is not public. The original 30 training images were therefore randomly split into a training set of 20 images and a test set of 10. Based on the same hyperparameters and computing environment as the above experiments, we used the original sized images of the EM dataset to train the discovered architecture, the original U-Net, FC-Densenet, CE-Net, and CS$^2$-Net. Their performance was tested on the test set. The ACC, SE, SP, F1, and AUROC metrics were used for statistical comparison. Table \ref{tab:EM}, shows that the discovered architecture also outperformed the other models on all five metrics. Fig. \ref{fig:EM} confirms that the discovered architecture achieved superior results to the other models, especially in challenging cases with irregular shapes or tiny and ambiguous boundaries, thus indicating the discovered architecture’s strong generalization ability in the cell boundary segmentation task.

\begin{table}[htbp]
	\centering
	\caption{Comparisons on cell membrane (boundary) segmentation.}
	\scalebox{0.7}{\begin{tabular}{cccccc}
		\toprule[1.5pt]
		Models & ACC   & SE    & SP    & F1 & AUROC \\
		\midrule[1.5pt]
		U-Net & 0.9112 & 0.8179 & 0.9401 & 0.8097 & 0.9587 \\
		FC-Densenet & 0.9132 & 0.8291 & 0.9361 & 0.8167 & 0.9627 \\
		CE-Net & 0.9099 & 0.8253 & 0.9233 & 0.8161 & 0.9645 \\
		CS$^2$-Net & 0.9109 & 0.8263 & 0.9272 & 0.8163 & 0.9634 \\
		Genetic U-Net & \textbf{0.9171} & \textbf{0.8376} & \textbf{0.9446} & \textbf{0.8212} & \textbf{0.9666} \\
		\bottomrule[1.5pt]
	\end{tabular}}%
	\label{tab:EM}%
\end{table}%
\subsection{Future work}
\label{sec:limitations}
Even though the network architecture obtained by the proposed method is capable of segmenting vessels closer to the ground truth than other models, it may fail in some very challenging cases. To improve the performance of the proposed approach further, a better design of the search space will be investigated. We must also extend the application of the proposed method to other scenarios to further verify its efficacy and generality.

\section{Conclusion}
\label{sec:6}
In this study, a novel neural architecture search (NAS) for retinal vessel segmentation, named Genetic U-Net, is proposed based on the U-shaped encoder-decoder structure. The existing CNN methods for retinal vessel segmentation can make hardly any further improvements and usually have profuse architecture-based parameters. Genetic U-Net employs an improved GA to evolve a model that outperforms existing mainstream methods in retinal vessel segmentation using a condensed but flexible search space that limits the number of architecture-based parameters within a range of small values. Therefore, the resulting architecture achieves a significant reduction in the number of parameters and the computational cost. This indicates that it is feasible to design an exceptional network architecture with fewer parameters for vessel segmentation that would facilitate its deployment for clinical applications and overcome the insufficiency of training data. Furthermore, by analyzing the evolved results, we found that utilizing several effective operations and patterns in the model’s building blocks greatly enhanced the performance in vessel segmentation, which provides instrumental domain knowledge for use in future studies.

\section*{Acknowledgment}
This work was supported by the Key Lab of Digital Signal and Image Processing of Guangdong Province, by the Key Laboratory of Intelligent Manufacturing Technology (Shantou University), Ministry of Education, by the Science and Technology Planning Project of Guangdong Province of China under grants (180917144960530, 2019A050519008, 2019A050520001), by the State Key Lab of Digital Manufacturing Equipment \& Technology under grant DMETKF2019020.

\ifCLASSOPTIONcaptionsoff
  \newpage
\fi

\normalem
\bibliographystyle{IEEEtran}

\bibliography{genetic-unet}

\onecolumn
\section{Appendix}
\subsection{Visualization of Discovered Achitectures}

\begin{figure*}[htbp]
	\centering
	\includegraphics[width=18cm]{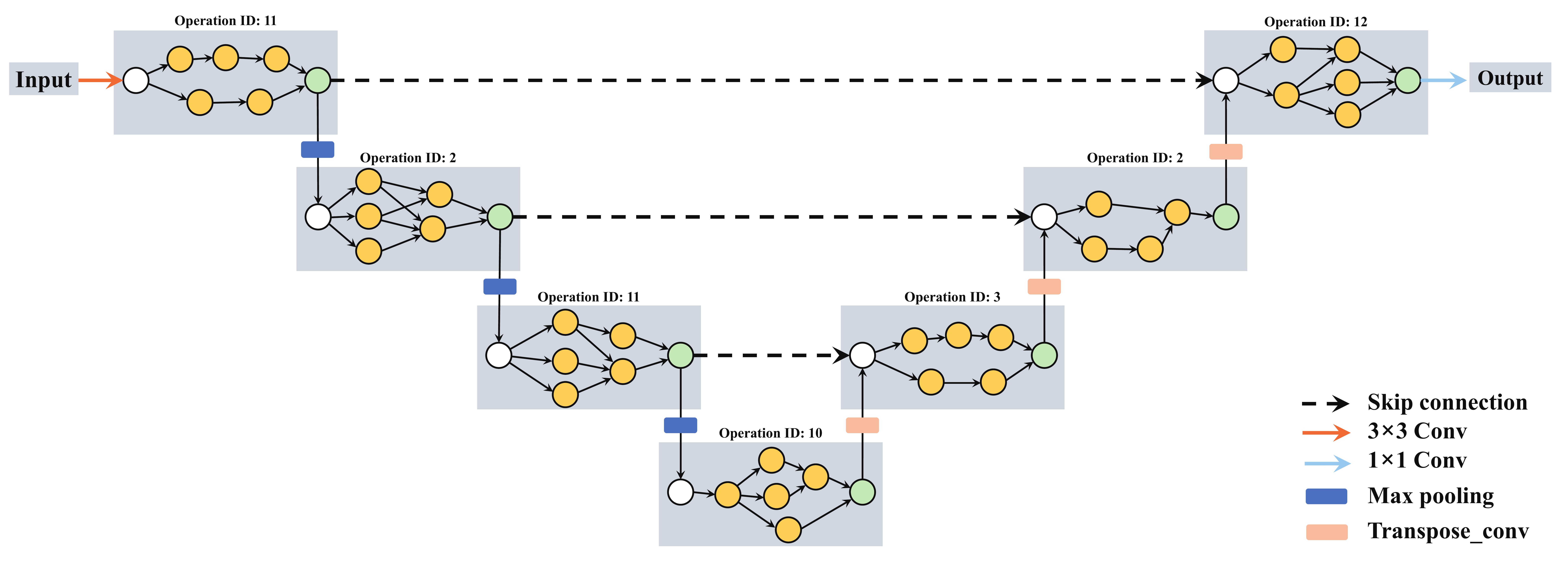}
	\caption{The second best architecture.}\label{fig:top2-arch}
	
\end{figure*}

\begin{figure*}[htbp]
	\centering
	\includegraphics[width=18cm]{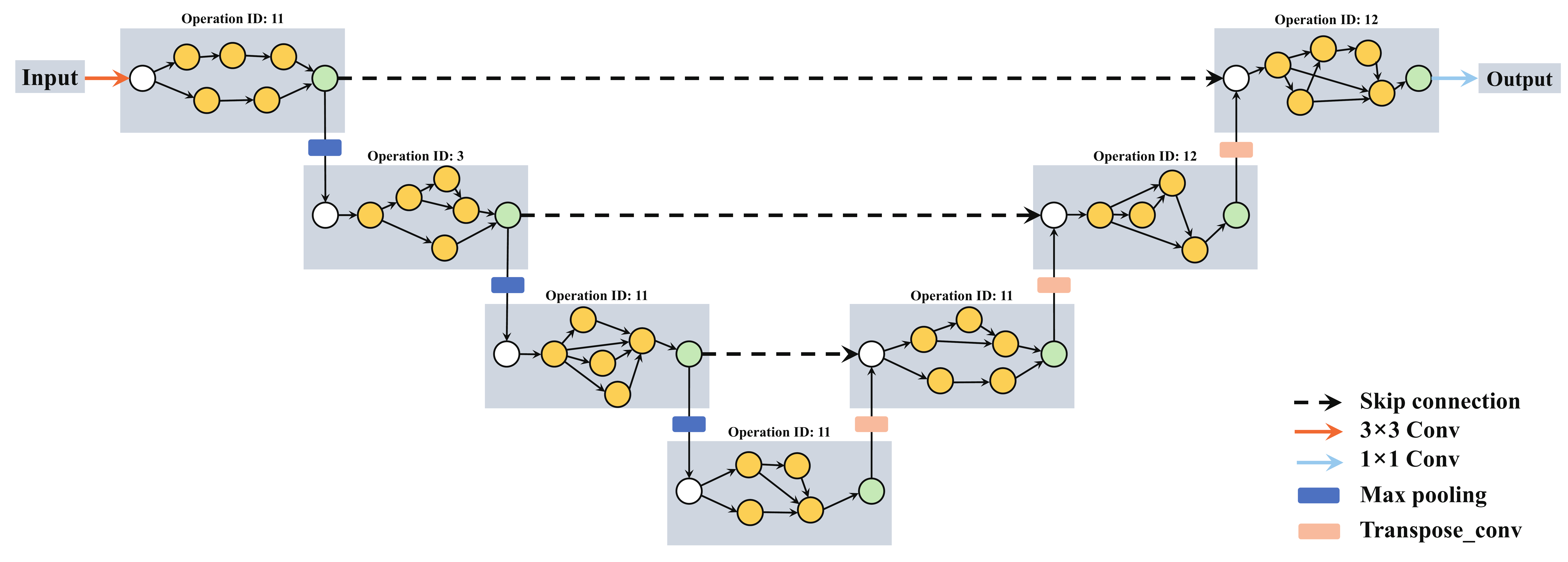}
	\caption{The third best architecture.}\label{fig:top3-arch}
	
\end{figure*}

\begin{figure*}[htbp]
	\centering
	\includegraphics[width=18cm]{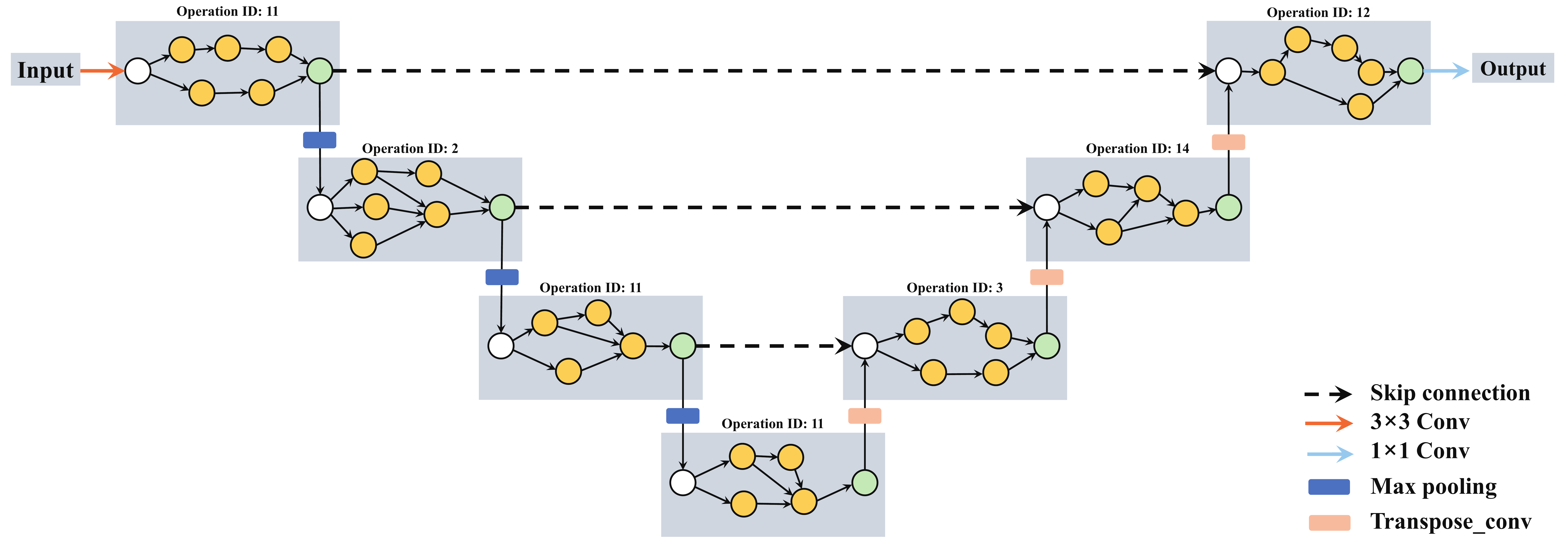}
	\caption{The fourth best architecture.}\label{fig:top4-arch}
	
\end{figure*}

\begin{figure*}[htbp]
	\centering
	\includegraphics[width=18cm]{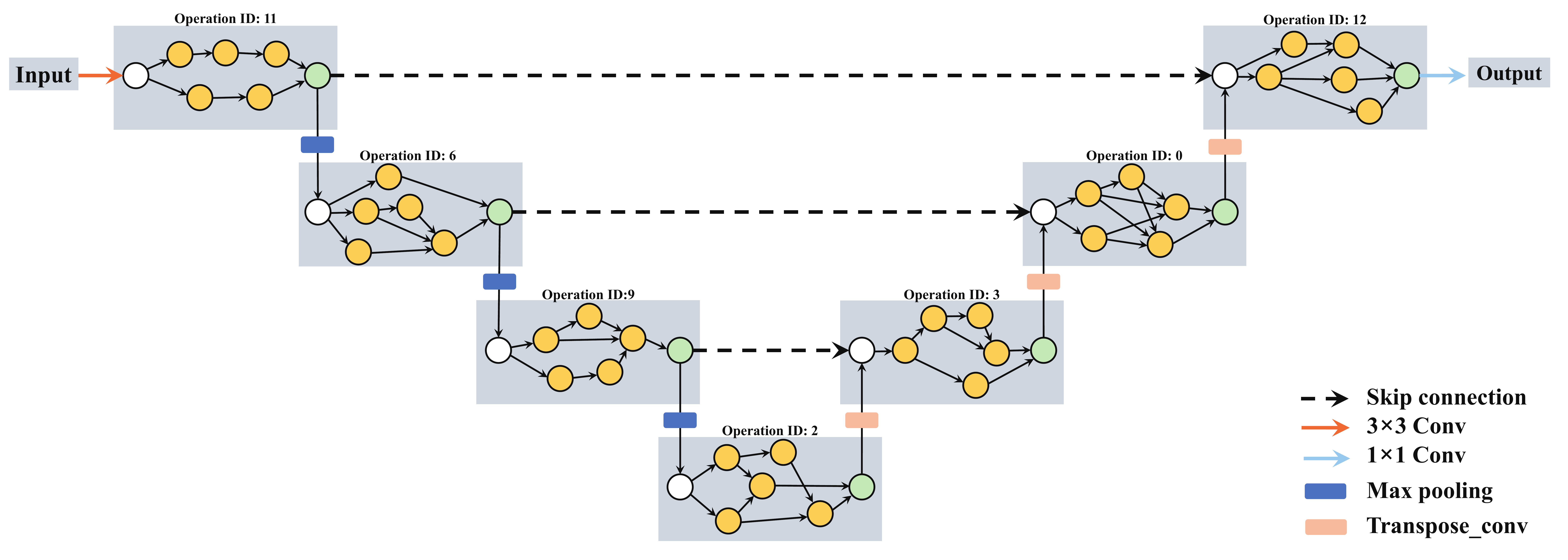}
	\caption{The fifth best architecture.}\label{fig:top5-arch}
	
\end{figure*}

\end{document}